\documentclass[prl, aps, showkeys, twocolumn, 10pt,
superscriptaddress, showpacs, floatfix
]{revtex4-1}
\usepackage{amsmath}
\usepackage{amssymb}
\usepackage[utf8]{inputenc}
\usepackage[english]{babel}
\usepackage{verbatim}
\usepackage{graphicx}
\usepackage{color}
\usepackage{hyphenat}
\usepackage{soul}
\usepackage[maxfloats=100]{morefloats}

\let\oldmarginpar\marginpar{}
\renewcommand\marginpar[1]{\oldmarginpar[\raggedleft\footnotesize #1]%
{\raggedright\footnotesize #1}}

\newcommand{\E}[1]{\langle{}#1 \rangle}

\newcommand{\lig}{\ensuremath{\mathrm{L}}{}}
\newcommand{\rec}{\ensuremath{\mathrm{R}}}
\newcommand{\lr}{\ensuremath{\mathrm{LR}}}
\newcommand{\nmax}{\ensuremath{N}}
\newcommand{\g}{\ensuremath{\tilde{\gamma}}}

\newcommand{\A}{{\ensuremath{\cal A}}}
\newcommand{\B}{\ensuremath{{\cal B}}}

\newcommand{\Em}{\mathcal{E}^-}
\newcommand{\Ep}{\mathcal{E}^+}
\newcommand{\Epm}{\mathcal{E}^\pm}
\newcommand{\hsl}{[s]}

\renewcommand{\r}[1]{\textcolor{red}{#1}}
\renewcommand{\r}[1]{#1}

\newcommand{\outline}[1]{}
\newcommand{\todo}[1]{}

\begin{document}


\author{Nils B. Becker}
\email{nbecker@bioquant.uni-heidelberg.de}
\affiliation{Bioquant, Universtit\"at Heidelberg, Im
Neuenheimer Feld 267, 69120 Heidelberg, Germany}

\author{Andrew Mugler}
\affiliation{Department of Physics, Purdue University, West Lafayette, IN
	47907, USA}

\author{Pieter Rein \surname{ten Wolde}}
\affiliation{FOM Institute AMOLF, Science Park 104, 1098 XG Amsterdam, The
Netherlands}

\title{Optimal Prediction by Cellular Signaling Networks}

\date{\today}

\begin{abstract}
Living cells can enhance their fitness by anticipating environmental change.
We study how accurately linear signaling networks in cells can predict future
signals.
We find that maximal predictive power results from a combination of input-noise
suppression, linear extrapolation, and selective readout of correlated
past signal values. Single-layer networks generate exponential response
kernels, which suffice to predict Markovian signals optimally. Multilayer
networks allow oscillatory kernels that can optimally predict non-Markovian
signals. At low noise, these kernels exploit the signal derivative for
extrapolation, while at high noise, they capitalize on signal values in
the past that are strongly correlated with the future signal. We show how the
common motifs of negative feedback and incoherent feed-forward can
implement these optimal response functions.
\r{Simulations reveal that \textit{E.\,coli} can reliably predict
  concentration changes for chemotaxis, and that the integration time
  of its response kernel arises from a trade-off between rapid
  response and noise suppression.}
\end{abstract}


\maketitle


The ability to respond and adapt to changing environments is a defining
property of life.  Single-celled organisms employ a range of response
strategies, tailored to the environmental fluctuations they encounter.
Gradual changes in osmolarity, pH or available nutrients are sensed and
responded to adiabatically. In this regime, the sensory performance as
measured by the mutual information between stimulus and response, limits the
achievable growth rate~\cite{bergstrom05,taylor07,bialek12}.
In contrast, when environmental changes are rapid and unpredictable, sensing
may be futile, since any response would come too late. Here, phenotypic
heterogeneity can help by providing a subpopulation of pre-adapted
cells~\cite{balaban04}.
An intermediate regime exists where environmental fluctuations occur with some
regularity, on the cellular response time scale. It is then possible and
desirable for the cell to \emph{predict} the future environment, in order to
initiate a response ahead of time. When the cellular response takes a finite
time $\tau$ to become effective, the \emph{predictive} mutual information
between the current sensory output and the environment $\tau$ later, limits
growth~\cite{supplementary}. Sensing strategies that leverage correlations
of a stimulus with future environmental changes have indeed been observed, and
re-evolved experimentally \cite{mitchell09,tagkopoulos08}.

This raises the question of what makes a cellular network an optimal
predictor, rather than instantaneous reporter, of the environment. Intuitively,
to predict, one should rely on the most up-to-date information,
\textit{i.e.}~respond to the current input.
However, cells often sense non-Markovian (NM) signals, whose past trajectories
could add useful information. Intriguingly, in such cases, sensory
networks often react not instantaneously but instead more slowly, on the time
scale of the signal~\cite{kaupp08,valencia12}.

A slow network time integrates the input signal, which may dampen the response,
but can also enhance the estimate of the current input signal by filtering
noise from, {\it e.g.}, receptor-ligand
binding~\cite{berg77,bialek05,govern12,mehta12a,kaizu14,govern14a}. Moreover, a
slow response may enhance prediction by building a memory of the
signal history which is informative about the future signal. What features of
signal and response then make a non-instantaneous response beneficial for
prediction?


Here, we study how the accuracy of prediction depends on the noise and
correlations in the input, the forecast interval, and the design of the
response system. We find that single-layer responders, such as push-pull
networks, can improve prediction by responding slowly. This not only allows
noise averaging, but also enables reading out past signals that are more
correlated with the future signal than the current signal is. Multilayer
networks can further enhance prediction via non-monotonic response functions
tailored to the input. They can optimally predict low-noise signals by
exploiting the signal derivative, and high-noise signals by coherently summing
informative past signal values. This can be imlemented via negative feedback.
\r{Finally, we perform simulations of \textit{E.\,coli} bacteria that chemotax
	in spatially varying concentration fields. The simulations reveal that
	\textit{E.\,coli} chemotaxis relies on predicting future concentration
changes. They suggest that the optimal integration time of the kernel arises as
a compromise between the benefit of responding quickly to the most recent
concentration values, and the need to filter input noise.}


\begin{figure}[tb]
	\begin{center}
		\includegraphics[width=1.\columnwidth]{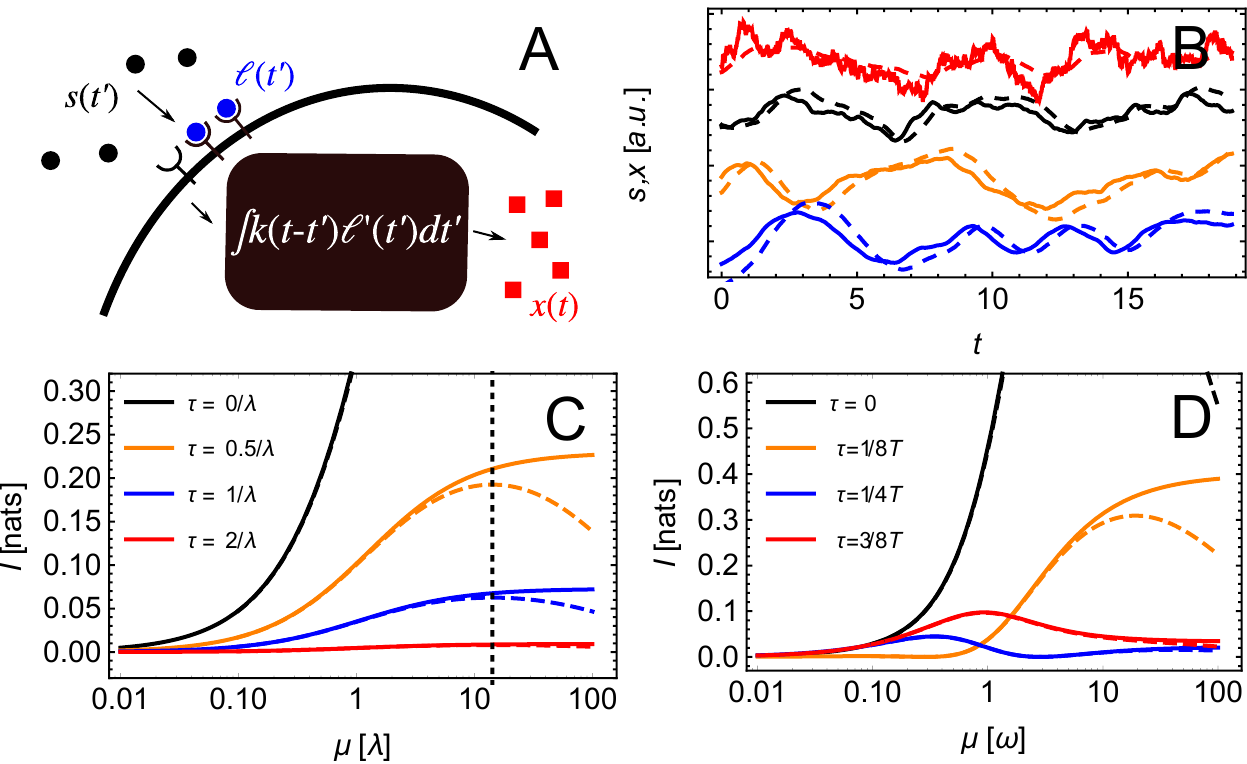}
		\caption{\label{fig:linearintro}%
			\r{%
			Biochemical prediction. (A) A sensory network (black box) with output $x$
			(red) senses an extracellular ligand $s$ (black) via noisy ligand-bound
			receptors $\ell$ (blue).
			(B) Example traces (solid lines) of Markovian
			($\lambda=1$, red) and non-Markovian (NM) signals $s$ ($\omega=1$,
			$\eta=4,2,0.5$, downwards), yielding outputs $x$ of exponential
			responders with $\mu=1$ (dashed).
			Predictive information $I[x,s_{\tau}]$ in nats ($\ln
			2~\mathrm{nats}=1~\mathrm{bit}$) for Markovian
			signals (C) and NM signals with damping $\eta=1/2$ (D),
			as a function of the speed $\mu$ of an exponential
			responder, for different prediction intervals $\tau$,
			and for noise level $\vartheta=0$ (solid lines) and
			$\vartheta=0.1$ (dashed lines).  Dotted line in (C)
			denotes $\mu_{\rm ti}$.}}
	\end{center}
\end{figure}

Consider a general sensory network which responds to a time-varying
extracellular signal by binding ligand molecules, relaying the signal via
intermediate species, and finally producing an output species
(Fig.~\ref{fig:linearintro}A). Its prediction capability depends on both
responder and input properties. Concerning the input, prediction fundamentally
requires that past inputs contain information about the future,
\textit{i.e.}~the signal's conditional probability density $p(s(t+\tau)|s(t),
s(t'),\dots)$ really depends on the signal values at $t>t'>\cdots$. For
Markovian input, the only dependence is on $s(t)$, and perfect instantaneous
readout of $s(t)$ would in fact be the optimal prediction strategy for all
future $s(t+\tau)$ \cite{supplementary}. However, in the presence of input
noise $\xi$, arising from, \emph{e.g}, receptor-ligand binding, the responder
senses the degraded signal $\ell(t)=s(t)+\xi(t)$. Then even for Markovian $s$,
the added noise makes $p(s(t+\tau)|\ell(t), \ell(t'), \dots)$ dependent on past
values $\ell(t'),\dots$, since they help determine the current input $s(t)$ by
averaging over the noise $\xi$, and then from $s(t)$ the future $s(t+\tau)$.
Thus a slow response can help prediction of any noisy signal via the mechanism
of time integration (which also improves accuracy for constant, noisy signals
\cite{berg77,bialek05,govern12,mehta12a,kaizu14,govern14a,govern14b,aquino14}).
As detailed below, for NM signals, another prediction mechanism exists: A
responder with memory enables readout of additional information from past
signals $s(t'),\dots$, improving predictions by exploiting signal correlations. 

We take the input signal $s(t)$ to be stationary Gaussian, characterized by
$\E{s(t')s(t'+t)}=\sigma_s^2 r_s(t)$ where $r_s$ denotes the normalized
autocorrelation function, and $\sigma_s$, the signal amplitude. For Markovian
processes, $r_s(t)=\exp(-\lambda t)$. \r{A family of NM signals} can be generated
via a harmonic oscillator defined by $\partial_{\omega t} q = p,\;
\partial_{\omega t}p = -q -\eta p + \sqrt{2\eta} \psi$ with unit white noise
$\psi$, \r{by letting $s\equiv q$, see Fig.~\ref{fig:linearintro}B}. The
damping parameter $\eta$ controls the signal statistics: in the overdamped
regime $\eta>2$, $r_s(t)$ is monotonically decreasing, while for $\eta<2$ it is
oscillatory with period approaching $T=2\pi/\omega$; \r{in both cases, the
signal $s$ obeys Gaussian statistics}. This family of signals allows analytical
results and interpolates from Markovian to non-Markovian, long-range
correlated, oscillatory signals. We model input noise as white,
$\E{\xi(t)\xi(t')} = \sigma_s^2\vartheta^2 \delta(t-t')$, where $\vartheta$ is
the relative noise strength.

Concerning the responder, we focus on linear signaling networks
\cite{heinrich02,govern12} which afford analytical results and
often describe information transmission remarkably
well~\cite{tanase-nicola06,ziv07,ronde10}. Since we are interested in
how prediction depends on the correlations and noise in the input, we consider
responders in the deterministic limit. The output $x(t)=\int_{-\infty}^t
k(t-t')\ell(t')dt'$ of the network is then determined by its linear response
function $k(t)$.

The predictive power of a signal-responder system is measured in a rigorous and
biologically relevant way~\cite{supplementary} by the predictive mutual
information $ I[x,s_\tau] =
		\big\langle
		\log\frac{p(x,s_\tau)}{p(x)p(s_\tau)}
		\big\rangle$
between the current output $x(t)$ and the future input $s_\tau\equiv
s(t+\tau)$. Since $x$ is jointly Gaussian with the input, the predictive
information reduces to a function $I[x,s_\tau] = -\frac12\log(1-r_{xs_\tau}^2)$
of the input-output correlation coefficient
\begin{align}\label{eq:iocorr}
	r_{xs_\tau} &= \frac{\Psi(\tau)}{[\Sigma+\Xi]^{1/2}}.
\end{align}
The overlap integral $\Psi (\tau) \equiv \int_0^\infty k(t)r_s(t+\tau)dt$ is
the part of the normalized output variance $\sigma^2_x/\sigma^2_s$ that is
correlated with the prediction target $s_\tau$. The denominator splits
$\sigma^2_{x}/\sigma^2_{s}$ into contributions from past signal,
$\Sigma\equiv \int_0^\infty k(t)r_s(t-t')k(t')dtdt'$, and past noise $\Xi\equiv
\vartheta^2 \int_0^\infty k(t)^2dt$~\cite{supplementary}.

We first consider a push-pull network, consisting of a single layer in
which the output $x$ is directly activated by the receptor. It is
characterized by an exponential kernel $k(t)\propto\exp(-\mu t)$ with response
speed $\mu$. Fig.~\ref{fig:linearintro}C shows how accurately such a
network can predict Markovian signals, as measured by the predictive
information $I$, \r{obtained analytically from
Eq.~\ref{eq:iocorr}~\cite{supplementary}}. Without input noise
($\vartheta\to0$), the fastest responders maximize the accuracy $I$, as
expected. When including input noise, there exists an optimal response speed
$\mu_\text{ti} = (2\lambda/\vartheta^2 + \lambda^2)^{1/2}$, independent of
$\tau$, and approaching $\mu_\text{ti} \to\lambda$ for high
noise~\cite{hinczewski14}. The optimum arises from a trade-off between rapid
tracking of the input and noise averaging~\cite{supplementary}.

Fig.~\ref{fig:linearintro}D shows $I$ for exponential responders predicting
oscillatory ($\eta=0.5$) NM signals. As before, input noise disfavors the
fastest responders. Interestingly, however, a finite response speed can be
optimal even when there is no input noise ($\vartheta=0$):
%
%
For prediction intervals above about a quarter period, frequency-matched
responders with $\mu_*\simeq\omega$ (obtained
numerically~\cite{supplementary}), perform best.

\begin{figure}[tb]
	\begin{center}
		\includegraphics[width=\columnwidth]{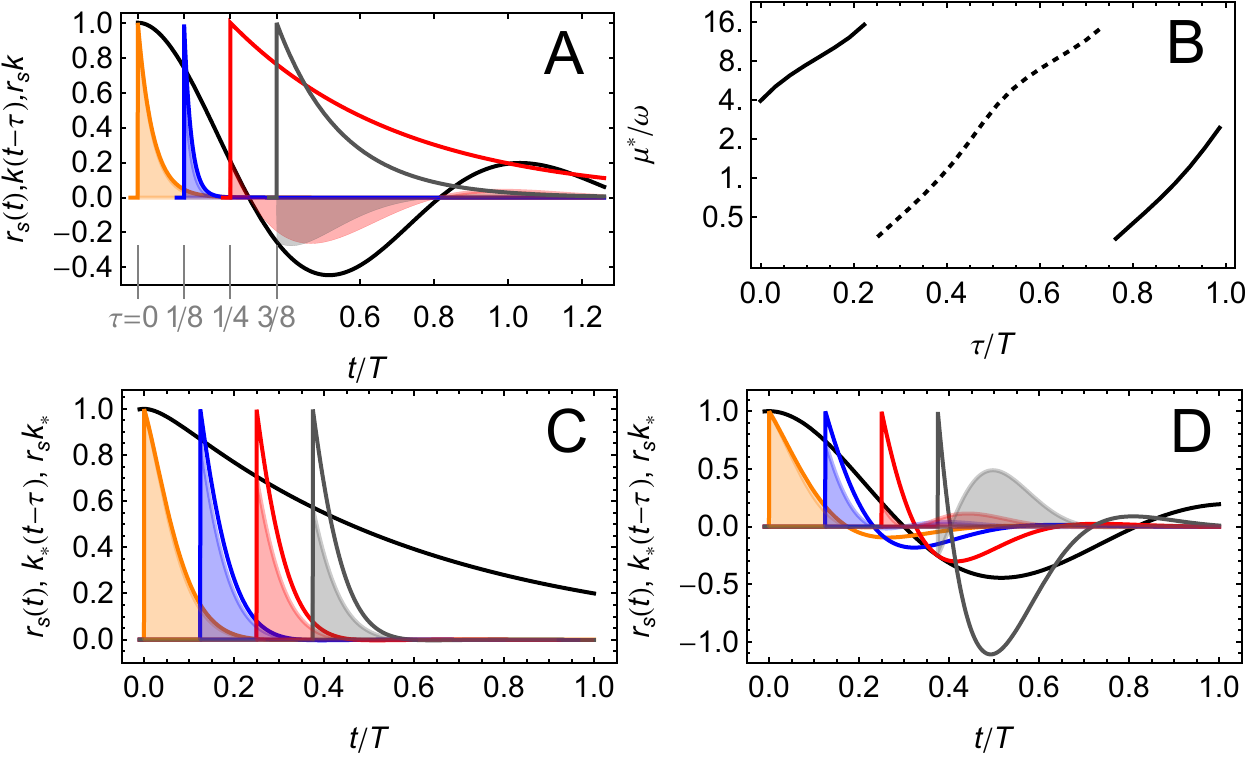}
		\caption{\label{fig:optkernels}%
			Prediction by optimizing correlations. (A) NM autocorrelation
			$r_s$ at $\eta=0.5$ (black). Optimally predictive exponential kernels
			$k(t-\tau)$ at noise level $\vartheta=0.25$ for $\tau/T=1/8,1/4,3/8$ as
			indicated (orange, blue, red, gray, respectively); with corresponding
			overlap integrands $r_s(t) k(t-\tau)$ (shaded). (B) Optimal $\mu_*$
			\textit{vs.}~prediction interval for these parameters. Solid, dashed
			lines: positive, negative $sx$ correlation, respectively. (C) As (A) at
			the same $\tau$ values but for the globally optimal kernels $k_*$ and
			overdamped signal $\eta=4$. (D) As (C) but for underdamped signal at
			$\eta=0.5$.}
	\end{center}
\end{figure}

The optimal $\mu_*$ is not an effect of simple time integration
but rather results from exploiting the oscillatory signal correlations.
When the forecast interval $\tau\ll T$, $r_{xs_\tau}^2$ is maximized by
increasing the overlap $\Psi(\tau)^2$ via a short kernel $k$ that samples high
values of the input correlation function $r_s$, Fig.~\ref{fig:optkernels}A,B.
The optimal kernels never become instantaneous, however, since that would
strongly increase $\Xi$.
As $\tau$ increases, $\mu_*$ initially increases: $k(t)$ decays faster so
that it continues to overlap with the positive lobe of $r_s(t+\tau)$; input
and output remain positively correlated.
Surprisingly, beyond a critical prediction interval $\tau_c\simeq 0.22T$,
$\mu_*$ drops discontinuously (Fig.~\ref{fig:optkernels}B, solid to
dashed line). The response now integrates the negative lobe of $r_s$,
anticorrelating output and input (Fig.~\ref{fig:optkernels}A). Effectively, the
output $x$ lags behind the input $s$ by an amount $\Lambda$, so that the
current output $x(t)$ reflects the past input $s(t-\Lambda)$ rather than the
current input $s(t)$. This enhances prediction, because the past signal
$s(t-\Lambda)$ is more (anti)correlated with, and hence more informative about,
the future $s(t+\tau)$ than the present signal $s(t)$ is, as shown by the
non-monotonic signal autocorrelation function: $r_s(\Lambda+\tau)^2 >
r_s(\tau)^2$ 
%
(cf.~Fig.~S2 in~\cite{supplementary}). 
The optimal response speed $\mu_*$ is such that $\tau+\Lambda \simeq T/2$; the
response kernel $k(t)$ then probes $r_s$ around its minimum, maximizing the
squared overlap $\Psi(\tau)^2$ between them (Fig.~\ref{fig:optkernels}A). As
$\tau$ increases further, increasing $\mu$ keeps the kernel localized in the
negative lobe of $r_s$, until another transition at higher $\tau\simeq 0.75T$
focuses the response on the next positive lobe of $r_s$. Simulations confirmed
this mechanism also for nonlinear responders and various input waveforms and
noise strengths~\cite{supplementary}.

Signaling networks typically consist of more than one layer~\cite{alon06},
generating complex kernels. To explore the design space, we maximize the
predictive information over all kernels. For Gaussian signals, this is
equivalent~\cite{supplementary} to finding the optimal kernel $k_*$ that
minimizes the mean squared prediction error $\E{(x-s_\tau)^2}$, as in
Wiener-\-Kol\-mo\-gorov filter theory
\cite{wiener50,kolmogorov92,hinczewski14,govern12}, used below.

The resulting optimal kernel remains exponential for input signals that are
Markovian~\cite{supplementary}, so that $k_*^\text{M}(t)\propto
\exp(-\mu_\text{ti} t)$, with $\mu_\text{ti}$, as before, implementing time
integration. Hence, a single, slowly responding, push-pull network
layer is enough to perform globally optimal predictions of noisy Markovian
signals; additional network layers cannot enhance prediction.

For NM but overdamped signals ($\eta>2$), optimal kernels $k_*$ have an
almost exponential shape, which is insensitive to the prediction interval,
Fig.~\ref{fig:optkernels}C (see~\cite{supplementary}). This indicates a
prediction strategy based mainly on time integration to determine the current
$s(t)$.

\begin{figure}[tb]
	\begin{center}
		\includegraphics[width=1.0\columnwidth]{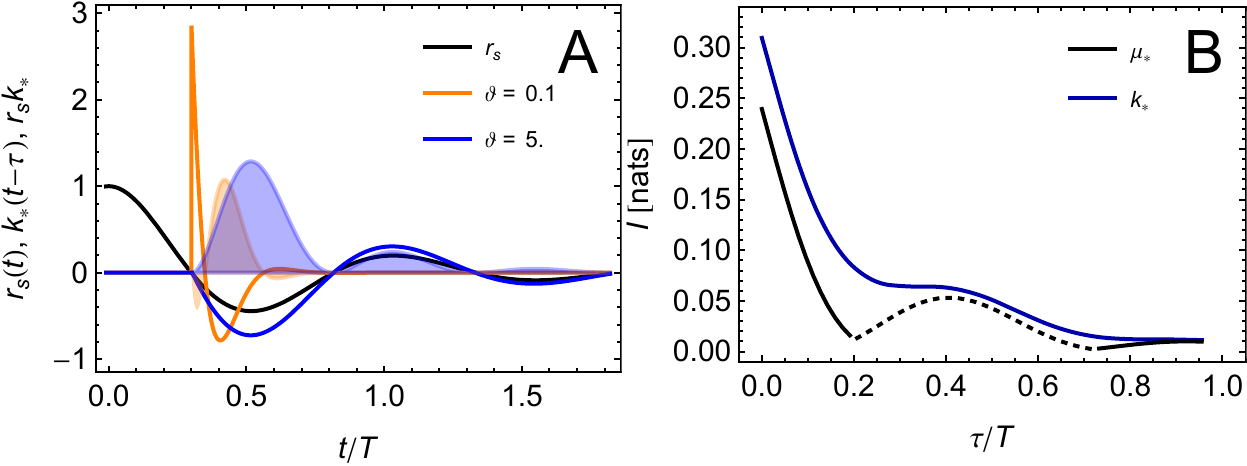}
		\caption{\label{fig:optkernelshighlow}%
			(A) The optimal kernel depends on the correlations and the
			noise $\vartheta$ in the input signal. Autocorrelation function (black)
			and optimal kernels $k_*$ at $\tau=0.3$ for oscillatory NM signals with
			$\eta=0.5$, for two different noise levels $\vartheta$. At low noise, the
			kernel consists of a positive lobe followed by a single undershoot. This
			corresponds to prediction based on linearly extrapolating the current
			signal. In contrast, at high noise, the kernel echoes the signal
			correlation function, exploiting signal values in the past that correlate
			with the future. (B) The optimal kernels $k_*$ strongly improve
			prediction over optimal exponential kernels $\mu_*$ (as in
			Fig.~\ref{fig:optkernels}B) around $\tau=\tau_c\approx 0.2T$;
			$\vartheta=1.0$; for low noise, see
			%
			%
			Fig.~S3~\cite{supplementary}.}
	\end{center}
\end{figure}

In contrast, oscillatory NM signals with $\eta<2$ yield optimal kernels that
are oscillatory, Fig.~\ref{fig:optkernels}D. Their shape depends on the
prediction interval $\tau$, and on the correlations and noise in the input
(Fig.~\ref{fig:optkernelshighlow}A). At low noise, optimal kernels integrate
only a short time window. They consist of a sharply-peaked positive lobe
followed by an undershoot, effectively estimating the future signal value from
its current value and derivative~\cite{supplementary}. This strategy of linear
extrapolation avoids including past signals, which are inherently less
correlated with the future. The capability to take derivatives enables a rapid
response even when the current signal value carries no predictive information,
$r_s(\tau)=0$; in contrast, in this situation exponential responders would need
to respond slowly, to pick up past, informative signals
(Fig.~\ref{fig:optkernels}).

As noise levels $\vartheta$ rise, noise averaging becomes increasingly
important, which demands longer kernels. However, to avoid signal damping, the
optimal kernel must \emph{coherently} sum past signal values. For oscillatory
signals, this requires an oscillatory kernel, which integrates the signal with
alternating signs. The prediction enhancement of globally optimal kernels over
optimal exponential kernels is indeed largest for oscillatory input signals
and, for $\tau\approx \tau_c$, it can reach up to 400\%
(Fig.~\ref{fig:optkernelshighlow}B). Interestingly, in the limit
$\vartheta\to\infty$, maximizing Eq.~\ref{eq:iocorr} gives the simple result
$k_*(t)\propto r_s(t+\tau)$~\cite{supplementary}, showing that at high noise,
the optimal kernel mimics the input correlations.

Optimal oscillatory kernel shapes like $k_*$ in
Fig.~\ref{fig:optkernelshighlow} can be implemented via negative feedback
\cite{supplementary}, a common motif in gene networks and signaling pathways
\cite{kondo97,kholodenko00,sasagawa05}. Another common motif, incoherent
feedforward~\cite{alon06}, only allows kernels with a positive lobe followed by
a single undershoot~\cite{supplementary}. Our results show that this is
useful for predicting low-noise non-Markovian signals, but suboptimal at high
noise.


In summary, accurate prediction requires capitalizing on past signal features
that are correlated with the future signal, while minimizing transmission of
uncorrelated past signals and noise. 
Single-layer networks suffice to predict Markovian signals optimally by noise
averaging. 
%
Multilayer networks predict oscillatory signals optimally, by fast linear
extrapolation at low noise, and by coherent summation at high noise. 
In the high noise limit, the optimal network response mimics the input: $k_*(t)
\propto r_s (\tau+t)$.

{To explore the importance of predictive power in cellular
	behavior, we have studied \textit{E.\,coli} chemotaxis. \textit{E.\,coli}
	moves by alternating straight \emph{runs} with \emph{tumbles}, which randomly
	reorient it. In a spatially varying environment, this motion is biased via a
	signaling pathway, whose output $x(t)$ controls the propensity $\alpha(t)$
	that a running bacterium will tumble. We have performed simulations of
	chemotaxing bacteria in static concentration fields $c(\vec r)$ in two
	dimensions, using the measured response kernel
	$k$~\cite{segall86,block82,celani10}.
At low concentrations, the signaling noise is dominated by the input noise. As
in our theory, we therefore ask how the predictive power depends on the kernel
and the input noise, ignoring intrinsic noise~\cite{berg04}. The tumbling
propensity is then given by $\alpha(t) = \alpha_0 [1-x(t)]$, where
$\alpha_0=1/\mathrm s$ is the basal tumbling rate and  $x(t) = \int_{-\infty}^t
k(t-t^\prime) \ell(t^\prime)dt^\prime$. The input $\ell(t')=s(t')+\xi(t')$
depends on the concentration signal $s(t)=c[\vec r(t)]$ and the input noise
$\xi(t)$ of relative strength $\theta$, arising \textit{e.g.}~from
receptor-ligand binding or receptor conformational dynamics. The kernel $k(t)$
is adaptive, \textit{i.e.}~integrates to 0, which allows the bacterium to
respond to a wide range of background concentrations \cite{segall86,block82}.
We compare adaptive kernels of varying range defined by $k_\nu(t) \equiv \nu^2
k(\nu t)$, where $\nu$ defines the response speed (see
also~\cite{supplementary}).

The sensory output modulates the \emph{delay} $\sim 1/\alpha(t)$ to the next
tumble. This suggests that high chemotactic efficiency requires accurate signal
prediction. However, it is less obvious what feature of the signal the system
actually predicts: The future concentration? Or the change in concentration?
More generally, what are the relevant input and output variables that control
chemotaxis? Only for these variables can we expect that chemotactic performance
is correlated with predictive information.

\begin{figure}[tb]
	\begin{center}
		\includegraphics[width=1.0\columnwidth]{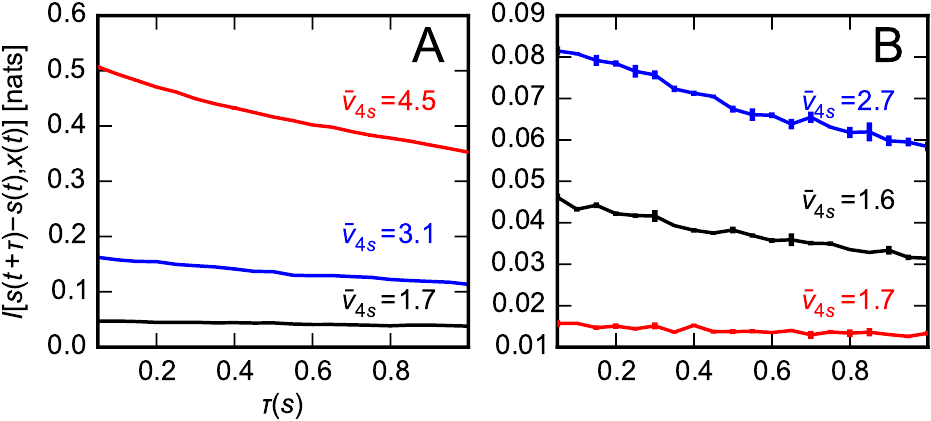}
		\caption{\label{fig:ivstau}%
			The predictive power of \textit{E.\,coli} in a sinusoidal concentration
			field with period $L=400\mu \mathrm{m}$, generating a
			non-oscillatory input signal~\cite{supplementary}, as a
			function of the forecast interval $\tau$. Information is shown for the
			wild-type \textit{E.\,coli} kernel with $\nu=1$ (blue), for a faster
			kernel with $\nu=3$ (red) and a slower kernel with $\nu=0.5$ (black), for
			noise levels $\theta=0$(A) and 2(B). Corresponding chemotactic
			speeds $\bar v_{\delta t=4\mathrm s}$ are given in $\mu \mathrm m/
			\mathrm s$. Bacteria run at $20\mu\mathrm m/\mathrm s$.}
	\end{center}
\end{figure}

To address this question, we performed simulations for three different kernels,
$\nu=0.5, 1, 3$ where $\nu=1$ corresponds to the measured kernel, and for two
input noise levels, $\theta=0, 2$. As our performance measure, we use the mean
chemotactic speed $\bar v_{\delta t} = \bigl\langle \frac{[\vec r(t+\delta
t)-\vec r(t)]\cdot \nabla c[\vec r(t)]}{\delta t \|\nabla c[\vec r(t)]\|}
\bigr\rangle$; similar results are obtained for the mean concentration
$\E{c[\vec r(t)]}$~\cite{supplementary}. We find that $\bar v_{\delta t}$ is
poorly correlated with the predictive information $I[x,s_\tau]$ between current
output and future concentration~\cite{supplementary}. In contrast, it is well
correlated with the predictive information $I[x, s_\tau - s]$ between current
output and future concentration \emph{change}, as Fig.~\ref{fig:ivstau} shows.
Hence, the search strategy of \textit{E.\,coli} is not based on predicting the
future concentration, but rather its trend, in accordance with the observation
that the bilobed kernel $k$ takes a time-derivative of the signal. If this is
positive, \textit{E.\,coli} `expects' that the concentration will continue to
rise, and will extend its run.

Fig.~\ref{fig:ivstau} also shows that the optimal kernel that maximizes the
information and hence chemotactic speed, depends on the input noise $\theta$. A
fast kernel emphasizes up-to-date information about recent concentration
changes, enabling an accurate and rapid response at low noise. At high noise,
its performance drops because it cannot filter the input noise and
hence cannot reliably predict future concentration changes. The optimal kernel
range then arises from a trade-off between agility and robustness.

Lastly, how far must \textit{E.\,coli} look into the future for efficient
chemotaxis? To anticipate concentration changes, the prediction horizon,
\textit{i.e.}~the time over which predictive information extends, should exceed
the response time. According to Eq.~\ref{eq:iocorr} the prediction horizon is
bounded by the signal correlation time, which is determined by the length
scale of the concentration field and by the motile behavior, to be explored in
future work. Already, Fig.~\ref{fig:ivstau} indicates that the prediction
horizon of \textit{E.\,coli} is indeed longer than the
response time, as $I(\tau)$ decays slower than $1/\alpha_0 = 1\mathrm s$. Our
results thus suggest that \textit{E.\,coli} can indeed reliably anticipate
concentration changes.

}



We thank Thomas Ouldridge, Tom Shimizu, and Giulia Malaguti for stimulating
discussions and a critical reading of the manuscript. This work is part of the
research program of the ``Stichting FOM'', which is financially supported by
the ``Nederlandse organisatie voor Wetenschappelijk Onderzoek'' (NWO). NBB was
supported by a fellowship of the Heidelberg Center for Modeling and Simulation
in the Biosciences (BIOMS).

\bibliography{draftliterature}

\begin{thebibliography}{47}%
\makeatletter
\providecommand \@ifxundefined [1]{%
 \@ifx{#1\undefined}
}%
\providecommand \@ifnum [1]{%
 \ifnum #1\expandafter \@firstoftwo
 \else \expandafter \@secondoftwo
 \fi
}%
\providecommand \@ifx [1]{%
 \ifx #1\expandafter \@firstoftwo
 \else \expandafter \@secondoftwo
 \fi
}%
\providecommand \natexlab [1]{#1}%
\providecommand \enquote  [1]{``#1''}%
\providecommand \bibnamefont  [1]{#1}%
\providecommand \bibfnamefont [1]{#1}%
\providecommand \citenamefont [1]{#1}%
\providecommand \href@noop [0]{\@secondoftwo}%
\providecommand \href [0]{\begingroup \@sanitize@url \@href}%
\providecommand \@href[1]{\@@startlink{#1}\@@href}%
\providecommand \@@href[1]{\endgroup#1\@@endlink}%
\providecommand \@sanitize@url [0]{\catcode `\\12\catcode `\$12\catcode
  `\&12\catcode `\#12\catcode `\^12\catcode `\_12\catcode `\%12\relax}%
\providecommand \@@startlink[1]{}%
\providecommand \@@endlink[0]{}%
\providecommand \url  [0]{\begingroup\@sanitize@url \@url }%
\providecommand \@url [1]{\endgroup\@href {#1}{\urlprefix }}%
\providecommand \urlprefix  [0]{URL }%
\providecommand \Eprint [0]{\href }%
\providecommand \doibase [0]{http://dx.doi.org/}%
\providecommand \selectlanguage [0]{\@gobble}%
\providecommand \bibinfo  [0]{\@secondoftwo}%
\providecommand \bibfield  [0]{\@secondoftwo}%
\providecommand \translation [1]{[#1]}%
\providecommand \BibitemOpen [0]{}%
\providecommand \bibitemStop [0]{}%
\providecommand \bibitemNoStop [0]{.\EOS\space}%
\providecommand \EOS [0]{\spacefactor3000\relax}%
\providecommand \BibitemShut  [1]{\csname bibitem#1\endcsname}%
\let\auto@bib@innerbib\@empty
\bibitem [{\citenamefont {Bergstrom}\ and\ \citenamefont
  {Lachmann}(2005)}]{bergstrom05}%
  \BibitemOpen
  \bibfield  {author} {\bibinfo {author} {\bibfnamefont {C.~T.}\ \bibnamefont
  {Bergstrom}}\ and\ \bibinfo {author} {\bibfnamefont {M.}~\bibnamefont
  {Lachmann}},\ }\href {http://arxiv.org/abs/q-bio/0510007} {\enquote {\bibinfo
  {title} {The fitness value of information},}\ } (\bibinfo {year} {2005}),\
  \Eprint {http://arxiv.org/abs/0510007 [q-bio.PE]} {arXiv:0510007 [q-bio.PE]}
  \BibitemShut {NoStop}%
\bibitem [{\citenamefont {Taylor}\ \emph {et~al.}(2007)\citenamefont {Taylor},
  \citenamefont {Tishby},\ and\ \citenamefont {Bialek}}]{taylor07}%
  \BibitemOpen
  \bibfield  {author} {\bibinfo {author} {\bibfnamefont {S.~F.}\ \bibnamefont
  {Taylor}}, \bibinfo {author} {\bibfnamefont {N.}~\bibnamefont {Tishby}}, \
  and\ \bibinfo {author} {\bibfnamefont {W.}~\bibnamefont {Bialek}},\
  }\href@noop {} {\enquote {\bibinfo {title} {Information and fitness},}\ }
  (\bibinfo {year} {2007}),\ \Eprint {http://arxiv.org/abs/0712.4382
  [q-bio.PE]} {arXiv:0712.4382 [q-bio.PE]} \BibitemShut {NoStop}%
\bibitem [{\citenamefont {Bialek}(2012)}]{bialek12}%
  \BibitemOpen
  \bibfield  {author} {\bibinfo {author} {\bibfnamefont {W.}~\bibnamefont
  {Bialek}},\ }\href@noop {} {\emph {\bibinfo {title} {Biophysics: Searching
  for Principles}}}\ (\bibinfo  {publisher} {Princeton Univ Pr},\ \bibinfo
  {year} {2012})\BibitemShut {NoStop}%
\bibitem [{\citenamefont {Balaban}\ \emph {et~al.}(2004)\citenamefont
  {Balaban}, \citenamefont {Merrin}, \citenamefont {Chait}, \citenamefont
  {Kowalik},\ and\ \citenamefont {Leibler}}]{balaban04}%
  \BibitemOpen
  \bibfield  {author} {\bibinfo {author} {\bibfnamefont {N.~Q.}\ \bibnamefont
  {Balaban}}, \bibinfo {author} {\bibfnamefont {J.}~\bibnamefont {Merrin}},
  \bibinfo {author} {\bibfnamefont {R.}~\bibnamefont {Chait}}, \bibinfo
  {author} {\bibfnamefont {L.}~\bibnamefont {Kowalik}}, \ and\ \bibinfo
  {author} {\bibfnamefont {S.}~\bibnamefont {Leibler}},\ }\href@noop {}
  {\bibfield  {journal} {\bibinfo  {journal} {Science}\ }\textbf {\bibinfo
  {volume} {305}},\ \bibinfo {pages} {1622} (\bibinfo {year}
  {2004})}\BibitemShut {NoStop}%
\bibitem [{sup()}]{supplementary}%
  \BibitemOpen
  \href@noop {} {\ }\bibinfo {note} {See supplementary
  information.}\BibitemShut {Stop}%
\bibitem [{\citenamefont {Mitchell}\ \emph {et~al.}(2009)\citenamefont
  {Mitchell}, \citenamefont {Romano}, \citenamefont {Groisman}, \citenamefont
  {Yona}, \citenamefont {Dekel}, \citenamefont {Kupiec}, \citenamefont
  {Dahan},\ and\ \citenamefont {Pilpel}}]{mitchell09}%
  \BibitemOpen
  \bibfield  {author} {\bibinfo {author} {\bibfnamefont {A.}~\bibnamefont
  {Mitchell}}, \bibinfo {author} {\bibfnamefont {G.~H.}\ \bibnamefont
  {Romano}}, \bibinfo {author} {\bibfnamefont {B.}~\bibnamefont {Groisman}},
  \bibinfo {author} {\bibfnamefont {A.}~\bibnamefont {Yona}}, \bibinfo {author}
  {\bibfnamefont {E.}~\bibnamefont {Dekel}}, \bibinfo {author} {\bibfnamefont
  {M.}~\bibnamefont {Kupiec}}, \bibinfo {author} {\bibfnamefont
  {O.}~\bibnamefont {Dahan}}, \ and\ \bibinfo {author} {\bibfnamefont
  {Y.}~\bibnamefont {Pilpel}},\ }\href {\doibase 10.1038/nature08112}
  {\bibfield  {journal} {\bibinfo  {journal} {Nature}\ }\textbf {\bibinfo
  {volume} {460}},\ \bibinfo {pages} {220} (\bibinfo {year}
  {2009})}\BibitemShut {NoStop}%
\bibitem [{\citenamefont {Tagkopoulos}\ \emph {et~al.}(2008)\citenamefont
  {Tagkopoulos}, \citenamefont {Liu},\ and\ \citenamefont
  {Tavazoie}}]{tagkopoulos08}%
  \BibitemOpen
  \bibfield  {author} {\bibinfo {author} {\bibfnamefont {I.}~\bibnamefont
  {Tagkopoulos}}, \bibinfo {author} {\bibfnamefont {Y.-C.}\ \bibnamefont
  {Liu}}, \ and\ \bibinfo {author} {\bibfnamefont {S.}~\bibnamefont
  {Tavazoie}},\ }\href {\doibase 10.1126/science.1154456} {\bibfield  {journal}
  {\bibinfo  {journal} {Science}\ }\textbf {\bibinfo {volume} {320}},\ \bibinfo
  {pages} {1313} (\bibinfo {year} {2008})},\ \bibinfo {note} {{PMID:}
  18467556}\BibitemShut {NoStop}%
\bibitem [{\citenamefont {Kaupp}\ \emph {et~al.}(2008)\citenamefont {Kaupp},
  \citenamefont {Kashikar},\ and\ \citenamefont {Weyand}}]{kaupp08}%
  \BibitemOpen
  \bibfield  {author} {\bibinfo {author} {\bibfnamefont {U.~B.}\ \bibnamefont
  {Kaupp}}, \bibinfo {author} {\bibfnamefont {N.~D.}\ \bibnamefont {Kashikar}},
  \ and\ \bibinfo {author} {\bibfnamefont {I.}~\bibnamefont {Weyand}},\ }\href
  {\doibase 10.1146/annurev.physiol.70.113006.100654} {\bibfield  {journal}
  {\bibinfo  {journal} {Annual Review of Physiology}\ }\textbf {\bibinfo
  {volume} {70}},\ \bibinfo {pages} {93} (\bibinfo {year} {2008})}\BibitemShut
  {NoStop}%
\bibitem [{\citenamefont {Valencia}\ \emph {et~al.}(2012)\citenamefont
  {Valencia}, \citenamefont {Bitou}, \citenamefont {Ishii}, \citenamefont
  {Murakami}, \citenamefont {Morishita}, \citenamefont {Onai}, \citenamefont
  {Furukawa}, \citenamefont {Imada}, \citenamefont {Namba},\ and\ \citenamefont
  {Ishiura}}]{valencia12}%
  \BibitemOpen
  \bibfield  {author} {\bibinfo {author} {\bibfnamefont {J.}~\bibnamefont
  {Valencia}, \bibfnamefont {S.}}, \bibinfo {author} {\bibfnamefont
  {K.}~\bibnamefont {Bitou}}, \bibinfo {author} {\bibfnamefont
  {K.}~\bibnamefont {Ishii}}, \bibinfo {author} {\bibfnamefont
  {R.}~\bibnamefont {Murakami}}, \bibinfo {author} {\bibfnamefont
  {M.}~\bibnamefont {Morishita}}, \bibinfo {author} {\bibfnamefont
  {K.}~\bibnamefont {Onai}}, \bibinfo {author} {\bibfnamefont {Y.}~\bibnamefont
  {Furukawa}}, \bibinfo {author} {\bibfnamefont {K.}~\bibnamefont {Imada}},
  \bibinfo {author} {\bibfnamefont {K.}~\bibnamefont {Namba}}, \ and\ \bibinfo
  {author} {\bibfnamefont {M.}~\bibnamefont {Ishiura}},\ }\href {\doibase
  10.1111/j.1365-2443.2012.01597.x} {\bibfield  {journal} {\bibinfo  {journal}
  {Genes to Cells}\ }\textbf {\bibinfo {volume} {17}},\ \bibinfo {pages} {398}
  (\bibinfo {year} {2012})}\BibitemShut {NoStop}%
\bibitem [{\citenamefont {Berg}\ and\ \citenamefont {Purcell}(1977)}]{berg77}%
  \BibitemOpen
  \bibfield  {author} {\bibinfo {author} {\bibfnamefont {H.~C.}\ \bibnamefont
  {Berg}}\ and\ \bibinfo {author} {\bibfnamefont {E.~M.}\ \bibnamefont
  {Purcell}},\ }\href {\doibase 10.1016/S0006-3495(77)85544-6} {\bibfield
  {journal} {\bibinfo  {journal} {Biophysical Journal}\ }\textbf {\bibinfo
  {volume} {20}},\ \bibinfo {pages} {193} (\bibinfo {year} {1977})}\BibitemShut
  {NoStop}%
\bibitem [{\citenamefont {Bialek}\ and\ \citenamefont
  {Setayeshgar}(2005)}]{bialek05}%
  \BibitemOpen
  \bibfield  {author} {\bibinfo {author} {\bibfnamefont {W.}~\bibnamefont
  {Bialek}}\ and\ \bibinfo {author} {\bibfnamefont {S.}~\bibnamefont
  {Setayeshgar}},\ }\href {\doibase 10.1073/pnas.0504321102} {\bibfield
  {journal} {\bibinfo  {journal} {Proceedings of the National Academy of
  Sciences of the United States of America}\ }\textbf {\bibinfo {volume}
  {102}},\ \bibinfo {pages} {10040} (\bibinfo {year} {2005})},\ \bibinfo {note}
  {{PMID:} 16006514}\BibitemShut {NoStop}%
\bibitem [{\citenamefont {Govern}\ and\ \citenamefont {ten
  Wolde}(2012)}]{govern12}%
  \BibitemOpen
  \bibfield  {author} {\bibinfo {author} {\bibfnamefont {C.~C.}\ \bibnamefont
  {Govern}}\ and\ \bibinfo {author} {\bibfnamefont {P.~R.}\ \bibnamefont {ten
  Wolde}},\ }\href {\doibase 10.1103/PhysRevLett.109.218103} {\bibfield
  {journal} {\bibinfo  {journal} {Physical Review Letters}\ }\textbf {\bibinfo
  {volume} {109}},\ \bibinfo {pages} {218103} (\bibinfo {year}
  {2012})}\BibitemShut {NoStop}%
\bibitem [{\citenamefont {Mehta}\ and\ \citenamefont
  {Schwab}(2012)}]{mehta12a}%
  \BibitemOpen
  \bibfield  {author} {\bibinfo {author} {\bibfnamefont {P.}~\bibnamefont
  {Mehta}}\ and\ \bibinfo {author} {\bibfnamefont {D.~J.}\ \bibnamefont
  {Schwab}},\ }\href {\doibase 10.1073/pnas.1207814109} {\bibfield  {journal}
  {\bibinfo  {journal} {Proceedings of the National Academy of Sciences}\ }
  (\bibinfo {year} {2012}),\ 10.1073/pnas.1207814109}\BibitemShut {NoStop}%
\bibitem [{\citenamefont {Kaizu}\ \emph {et~al.}(2014)\citenamefont {Kaizu},
  \citenamefont {de~Ronde}, \citenamefont {Paijmans}, \citenamefont
  {Takahashi}, \citenamefont {Tostevin},\ and\ \citenamefont {ten
  Wolde}}]{kaizu14}%
  \BibitemOpen
  \bibfield  {author} {\bibinfo {author} {\bibfnamefont {K.}~\bibnamefont
  {Kaizu}}, \bibinfo {author} {\bibfnamefont {W.}~\bibnamefont {de~Ronde}},
  \bibinfo {author} {\bibfnamefont {J.}~\bibnamefont {Paijmans}}, \bibinfo
  {author} {\bibfnamefont {K.}~\bibnamefont {Takahashi}}, \bibinfo {author}
  {\bibfnamefont {F.}~\bibnamefont {Tostevin}}, \ and\ \bibinfo {author}
  {\bibfnamefont {P.~R.}\ \bibnamefont {ten Wolde}},\ }\href {\doibase
  10.1016/j.bpj.2013.12.030} {\bibfield  {journal} {\bibinfo  {journal}
  {Biophysical Journal}\ }\textbf {\bibinfo {volume} {106}},\ \bibinfo {pages}
  {976} (\bibinfo {year} {2014})}\BibitemShut {NoStop}%
\bibitem [{\citenamefont {Govern}\ and\ \citenamefont {ten
  Wolde}(2014)}]{govern14a}%
  \BibitemOpen
  \bibfield  {author} {\bibinfo {author} {\bibfnamefont {C.~C.}\ \bibnamefont
  {Govern}}\ and\ \bibinfo {author} {\bibfnamefont {P.~R.}\ \bibnamefont {ten
  Wolde}},\ }\href {\doibase 10.1073} {\bibfield  {journal} {\bibinfo
  {journal} {Proceedings of the National Academy of Sciences of the United
  States of America}\ }\textbf {\bibinfo {volume} {111}},\ \bibinfo {pages}
  {17486} (\bibinfo {year} {2014})}\BibitemShut {NoStop}%
\bibitem [{\citenamefont {Govern}\ and\ \citenamefont {{ten
  Wolde}}(2014)}]{govern14b}%
  \BibitemOpen
  \bibfield  {author} {\bibinfo {author} {\bibfnamefont {C.~C.}\ \bibnamefont
  {Govern}}\ and\ \bibinfo {author} {\bibfnamefont {P.~R.}\ \bibnamefont {{ten
  Wolde}}},\ }\href@noop {} {\bibfield  {journal} {\bibinfo  {journal} {Phys
  Rev Lett}\ }\textbf {\bibinfo {volume} {113}},\ \bibinfo {pages} {258102}
  (\bibinfo {year} {2014})}\BibitemShut {NoStop}%
\bibitem [{\citenamefont {Aquino}\ \emph {et~al.}(2014)\citenamefont {Aquino},
  \citenamefont {Tweedy}, \citenamefont {Heinrich},\ and\ \citenamefont
  {Endres}}]{aquino14}%
  \BibitemOpen
  \bibfield  {author} {\bibinfo {author} {\bibfnamefont {G.}~\bibnamefont
  {Aquino}}, \bibinfo {author} {\bibfnamefont {L.}~\bibnamefont {Tweedy}},
  \bibinfo {author} {\bibfnamefont {D.}~\bibnamefont {Heinrich}}, \ and\
  \bibinfo {author} {\bibfnamefont {R.~G.}\ \bibnamefont {Endres}},\ }\href
  {\doibase 10.1038/srep05688} {\bibfield  {journal} {\bibinfo  {journal}
  {Scientific Reports}\ }\textbf {\bibinfo {volume} {4}} (\bibinfo {year}
  {2014}),\ 10.1038/srep05688}\BibitemShut {NoStop}%
\bibitem [{\citenamefont {Heinrich}\ \emph {et~al.}(2002)\citenamefont
  {Heinrich}, \citenamefont {Neel},\ and\ \citenamefont
  {Rapoport}}]{heinrich02}%
  \BibitemOpen
  \bibfield  {author} {\bibinfo {author} {\bibfnamefont {R.}~\bibnamefont
  {Heinrich}}, \bibinfo {author} {\bibfnamefont {B.~G.}\ \bibnamefont {Neel}},
  \ and\ \bibinfo {author} {\bibfnamefont {T.~A.}\ \bibnamefont {Rapoport}},\
  }\href {\doibase 10.1016/S1097-2765(02)00528-2} {\bibfield  {journal}
  {\bibinfo  {journal} {Molecular Cell}\ }\textbf {\bibinfo {volume} {9}},\
  \bibinfo {pages} {957} (\bibinfo {year} {2002})}\BibitemShut {NoStop}%
\bibitem [{\citenamefont {Tănase-Nicola}\ \emph {et~al.}(2006)\citenamefont
  {Tănase-Nicola}, \citenamefont {Warren},\ and\ \citenamefont {ten
  Wolde}}]{tanase-nicola06}%
  \BibitemOpen
  \bibfield  {author} {\bibinfo {author} {\bibfnamefont {S.}~\bibnamefont
  {Tănase-Nicola}}, \bibinfo {author} {\bibfnamefont {P.~B.}\ \bibnamefont
  {Warren}}, \ and\ \bibinfo {author} {\bibfnamefont {P.~R.}\ \bibnamefont {ten
  Wolde}},\ }\href {\doibase 10.1103/PhysRevLett.97.068102} {\bibfield
  {journal} {\bibinfo  {journal} {Physical Review Letters}\ }\textbf {\bibinfo
  {volume} {97}},\ \bibinfo {pages} {068102} (\bibinfo {year}
  {2006})}\BibitemShut {NoStop}%
\bibitem [{\citenamefont {Ziv}\ \emph {et~al.}(2007)\citenamefont {Ziv},
  \citenamefont {Nemenman},\ and\ \citenamefont {Wiggins}}]{ziv07}%
  \BibitemOpen
  \bibfield  {author} {\bibinfo {author} {\bibfnamefont {E.}~\bibnamefont
  {Ziv}}, \bibinfo {author} {\bibfnamefont {I.}~\bibnamefont {Nemenman}}, \
  and\ \bibinfo {author} {\bibfnamefont {C.~H.}\ \bibnamefont {Wiggins}},\
  }\href {\doibase 10.1371/journal.pone.0001077} {\bibfield  {journal}
  {\bibinfo  {journal} {PLoS One}\ }\textbf {\bibinfo {volume} {2}},\ \bibinfo
  {pages} {e1077} (\bibinfo {year} {2007})}\BibitemShut {NoStop}%
\bibitem [{\citenamefont {de~Ronde}\ \emph {et~al.}(2010)\citenamefont
  {de~Ronde}, \citenamefont {Tostevin},\ and\ \citenamefont {ten
  Wolde}}]{ronde10}%
  \BibitemOpen
  \bibfield  {author} {\bibinfo {author} {\bibfnamefont {W.~H.}\ \bibnamefont
  {de~Ronde}}, \bibinfo {author} {\bibfnamefont {F.}~\bibnamefont {Tostevin}},
  \ and\ \bibinfo {author} {\bibfnamefont {P.~R.}\ \bibnamefont {ten Wolde}},\
  }\href {\doibase 10.1103/PhysRevE.82.031914} {\bibfield  {journal} {\bibinfo
  {journal} {Physical Review E}\ }\textbf {\bibinfo {volume} {82}},\ \bibinfo
  {pages} {031914} (\bibinfo {year} {2010})}\BibitemShut {NoStop}%
\bibitem [{\citenamefont {Hinczewski}\ and\ \citenamefont
  {Thirumalai}(2014)}]{hinczewski14}%
  \BibitemOpen
  \bibfield  {author} {\bibinfo {author} {\bibfnamefont {M.}~\bibnamefont
  {Hinczewski}}\ and\ \bibinfo {author} {\bibfnamefont {D.}~\bibnamefont
  {Thirumalai}},\ }\href {\doibase 10.1103/PhysRevX.4.041017} {\bibfield
  {journal} {\bibinfo  {journal} {Physical Review X}\ }\textbf {\bibinfo
  {volume} {4}},\ \bibinfo {pages} {041017} (\bibinfo {year}
  {2014})}\BibitemShut {NoStop}%
\bibitem [{\citenamefont {Alon}(2006)}]{alon06}%
  \BibitemOpen
  \bibfield  {author} {\bibinfo {author} {\bibfnamefont {U.}~\bibnamefont
  {Alon}},\ }\href@noop {} {\emph {\bibinfo {title} {An Introduction to Systems
  Biology: Design Principles of Biological Circuits}}},\ \bibinfo {edition}
  {1st}\ ed.\ (\bibinfo  {publisher} {Crc Pr Inc},\ \bibinfo {year}
  {2006})\BibitemShut {NoStop}%
\bibitem [{\citenamefont {Wiener}(1950)}]{wiener50}%
  \BibitemOpen
  \bibfield  {author} {\bibinfo {author} {\bibfnamefont {N.}~\bibnamefont
  {Wiener}},\ }\href@noop {} {\emph {\bibinfo {title} {Extrapolation,
  interpolation, and smoothing of stationary time series}}}\ (\bibinfo
  {publisher} {Wiley},\ \bibinfo {year} {1950})\BibitemShut {NoStop}%
\bibitem [{\citenamefont {Kolmogorov}(1992)}]{kolmogorov92}%
  \BibitemOpen
  \bibfield  {author} {\bibinfo {author} {\bibfnamefont {A.~N.}\ \bibnamefont
  {Kolmogorov}},\ }\href@noop {} {\emph {\bibinfo {title} {Selected Works of A.
  N. Kolmogorov: Probability theory and mathematical statistics}}}\ (\bibinfo
  {publisher} {Springer Science \& Business Media},\ \bibinfo {year}
  {1992})\BibitemShut {NoStop}%
\bibitem [{\citenamefont {Kondo}(1997)}]{kondo97}%
  \BibitemOpen
  \bibfield  {author} {\bibinfo {author} {\bibfnamefont {T.}~\bibnamefont
  {Kondo}},\ }\href@noop {} {\bibfield  {journal} {\bibinfo  {journal}
  {Science}\ }\textbf {\bibinfo {volume} {275}},\ \bibinfo {pages} {224}
  (\bibinfo {year} {1997})}\BibitemShut {NoStop}%
\bibitem [{\citenamefont {Kholodenko}(2000)}]{kholodenko00}%
  \BibitemOpen
  \bibfield  {author} {\bibinfo {author} {\bibfnamefont {B.~N.}\ \bibnamefont
  {Kholodenko}},\ }\href {http://www.ncbi.nlm.nih.gov/pubmed/10712587}
  {\bibfield  {journal} {\bibinfo  {journal} {European journal of biochemistry
  / FEBS}\ }\textbf {\bibinfo {volume} {267}},\ \bibinfo {pages} {1583}
  (\bibinfo {year} {2000})}\BibitemShut {NoStop}%
\bibitem [{\citenamefont {Sasagawa}\ \emph {et~al.}(2005)\citenamefont
  {Sasagawa}, \citenamefont {Ozaki}, \citenamefont {Fujita},\ and\
  \citenamefont {Kuroda}}]{sasagawa05}%
  \BibitemOpen
  \bibfield  {author} {\bibinfo {author} {\bibfnamefont {S.}~\bibnamefont
  {Sasagawa}}, \bibinfo {author} {\bibfnamefont {Y.-i.}\ \bibnamefont {Ozaki}},
  \bibinfo {author} {\bibfnamefont {K.}~\bibnamefont {Fujita}}, \ and\ \bibinfo
  {author} {\bibfnamefont {S.}~\bibnamefont {Kuroda}},\ }\href {\doibase
  10.1038/ncb1233} {\bibfield  {journal} {\bibinfo  {journal} {Nat Cell Biol}\
  }\textbf {\bibinfo {volume} {7}},\ \bibinfo {pages} {365} (\bibinfo {year}
  {2005})}\BibitemShut {NoStop}%
\bibitem [{\citenamefont {Segall}\ \emph {et~al.}(1986)\citenamefont {Segall},
  \citenamefont {Block},\ and\ \citenamefont {Berg}}]{segall86}%
  \BibitemOpen
  \bibfield  {author} {\bibinfo {author} {\bibfnamefont {J.~E.}\ \bibnamefont
  {Segall}}, \bibinfo {author} {\bibfnamefont {S.~M.}\ \bibnamefont {Block}}, \
  and\ \bibinfo {author} {\bibfnamefont {H.~C.}\ \bibnamefont {Berg}},\
  }\href@noop {} {\bibfield  {journal} {\bibinfo  {journal} {Proc Natl Acad Sci
  U S A}\ }\textbf {\bibinfo {volume} {83}},\ \bibinfo {pages} {8987} (\bibinfo
  {year} {1986})}\BibitemShut {NoStop}%
\bibitem [{\citenamefont {Block}\ \emph {et~al.}(1982)\citenamefont {Block},
  \citenamefont {Segall},\ and\ \citenamefont {Berg}}]{block82}%
  \BibitemOpen
  \bibfield  {author} {\bibinfo {author} {\bibfnamefont {S.~M.}\ \bibnamefont
  {Block}}, \bibinfo {author} {\bibfnamefont {J.~E.}\ \bibnamefont {Segall}}, \
  and\ \bibinfo {author} {\bibfnamefont {H.~C.}\ \bibnamefont {Berg}},\
  }\href@noop {} {\bibfield  {journal} {\bibinfo  {journal} {Cell}\ }\textbf
  {\bibinfo {volume} {31}},\ \bibinfo {pages} {215} (\bibinfo {year}
  {1982})}\BibitemShut {NoStop}%
\bibitem [{\citenamefont {Celani}\ and\ \citenamefont
  {Vergassola}(2010)}]{celani10}%
  \BibitemOpen
  \bibfield  {author} {\bibinfo {author} {\bibfnamefont {A.}~\bibnamefont
  {Celani}}\ and\ \bibinfo {author} {\bibfnamefont {M.}~\bibnamefont
  {Vergassola}},\ }\href {\doibase 10.1073/pnas.0909673107} {\bibfield
  {journal} {\bibinfo  {journal} {Proc Natl Acad Sci U S A}\ }\textbf {\bibinfo
  {volume} {107}},\ \bibinfo {pages} {1391} (\bibinfo {year}
  {2010})}\BibitemShut {NoStop}%
\bibitem [{\citenamefont {Berg}(2004)}]{berg04}%
  \BibitemOpen
  \bibfield  {author} {\bibinfo {author} {\bibfnamefont {H.}~\bibnamefont
  {Berg}},\ }\href@noop {} {\emph {\bibinfo {title} {{E. $coli$ in motion}}}},\
  Biological and Medical Physics Biomedical Engineering\ (\bibinfo  {publisher}
  {Springer},\ \bibinfo {address} {New York},\ \bibinfo {year}
  {2004})\BibitemShut {NoStop}%
\bibitem [{\citenamefont {Bialek}\ \emph {et~al.}(2001)\citenamefont {Bialek},
  \citenamefont {Nemenman},\ and\ \citenamefont {Tishby}}]{bialek01}%
  \BibitemOpen
  \bibfield  {author} {\bibinfo {author} {\bibfnamefont {W.}~\bibnamefont
  {Bialek}}, \bibinfo {author} {\bibfnamefont {I.}~\bibnamefont {Nemenman}}, \
  and\ \bibinfo {author} {\bibfnamefont {N.}~\bibnamefont {Tishby}},\ }\href
  {\doibase 10.1162/089976601753195969} {\bibfield  {journal} {\bibinfo
  {journal} {Neural Comput}\ }\textbf {\bibinfo {volume} {13}},\ \bibinfo
  {pages} {2409} (\bibinfo {year} {2001})}\BibitemShut {NoStop}%
\bibitem [{\citenamefont {Tostevin}\ and\ \citenamefont {{ten
  Wolde}}(2009)}]{tostevin09}%
  \BibitemOpen
  \bibfield  {author} {\bibinfo {author} {\bibfnamefont {F.}~\bibnamefont
  {Tostevin}}\ and\ \bibinfo {author} {\bibfnamefont {P.~R.}\ \bibnamefont
  {{ten Wolde}}},\ }\href@noop {} {\bibfield  {journal} {\bibinfo  {journal}
  {Phys Rev Lett}\ }\textbf {\bibinfo {volume} {102}},\ \bibinfo {pages}
  {218101} (\bibinfo {year} {2009})}\BibitemShut {NoStop}%
\bibitem [{\citenamefont {Cover}\ and\ \citenamefont {Thomas}(2012)}]{cover12}%
  \BibitemOpen
  \bibfield  {author} {\bibinfo {author} {\bibfnamefont {T.~M.}\ \bibnamefont
  {Cover}}\ and\ \bibinfo {author} {\bibfnamefont {J.~A.}\ \bibnamefont
  {Thomas}},\ }\href@noop {} {\emph {\bibinfo {title} {Elements of information
  theory}}}\ (\bibinfo  {publisher} {John Wiley \& Sons},\ \bibinfo {year}
  {2012})\BibitemShut {NoStop}%
\bibitem [{\citenamefont {Tostevin}\ and\ \citenamefont {{ten
  Wolde}}(2010)}]{tostevin10}%
  \BibitemOpen
  \bibfield  {author} {\bibinfo {author} {\bibfnamefont {F.}~\bibnamefont
  {Tostevin}}\ and\ \bibinfo {author} {\bibfnamefont {P.~R.}\ \bibnamefont
  {{ten Wolde}}},\ }\href@noop {} {\bibfield  {journal} {\bibinfo  {journal}
  {Phys Rev E Stat Nonlin Soft Matter Phys}\ }\textbf {\bibinfo {volume}
  {81}},\ \bibinfo {pages} {061917} (\bibinfo {year} {2010})}\BibitemShut
  {NoStop}%
\bibitem [{\citenamefont {Mugler}\ \emph {et~al.}(2010)\citenamefont {Mugler},
  \citenamefont {Walczak},\ and\ \citenamefont {Wiggins}}]{Mugler2010}%
  \BibitemOpen
  \bibfield  {author} {\bibinfo {author} {\bibfnamefont {A.}~\bibnamefont
  {Mugler}}, \bibinfo {author} {\bibfnamefont {A.~M.}\ \bibnamefont {Walczak}},
  \ and\ \bibinfo {author} {\bibfnamefont {C.~H.}\ \bibnamefont {Wiggins}},\
  }\href {\doibase 10.1103/PhysRevLett.105.058101} {\bibfield  {journal}
  {\bibinfo  {journal} {Phys. Rev. Lett.}\ }\textbf {\bibinfo {volume} {105}},\
  \bibinfo {pages} {058101} (\bibinfo {year} {2010})}\BibitemShut {NoStop}%
\bibitem [{\citenamefont {Walczak}\ \emph {et~al.}(2009)\citenamefont
  {Walczak}, \citenamefont {Mugler},\ and\ \citenamefont
  {Wiggins}}]{Walczak2009}%
  \BibitemOpen
  \bibfield  {author} {\bibinfo {author} {\bibfnamefont {A.~M.}\ \bibnamefont
  {Walczak}}, \bibinfo {author} {\bibfnamefont {A.}~\bibnamefont {Mugler}}, \
  and\ \bibinfo {author} {\bibfnamefont {C.~H.}\ \bibnamefont {Wiggins}},\
  }\href {\doibase 10.1073/pnas.0811999106} {\bibfield  {journal} {\bibinfo
  {journal} {Proceedings of the National Academy of Sciences}\ }\textbf
  {\bibinfo {volume} {106}},\ \bibinfo {pages} {6529} (\bibinfo {year}
  {2009})}\BibitemShut {NoStop}%
\bibitem [{\citenamefont {Mugler}\ \emph {et~al.}(2009)\citenamefont {Mugler},
  \citenamefont {Walczak},\ and\ \citenamefont {Wiggins}}]{Mugler2009}%
  \BibitemOpen
  \bibfield  {author} {\bibinfo {author} {\bibfnamefont {A.}~\bibnamefont
  {Mugler}}, \bibinfo {author} {\bibfnamefont {A.~M.}\ \bibnamefont {Walczak}},
  \ and\ \bibinfo {author} {\bibfnamefont {C.~H.}\ \bibnamefont {Wiggins}},\
  }\href {\doibase 10.1103/PhysRevE.80.041921} {\bibfield  {journal} {\bibinfo
  {journal} {Phys. Rev. E}\ }\textbf {\bibinfo {volume} {80}},\ \bibinfo
  {pages} {41921} (\bibinfo {year} {2009})}\BibitemShut {NoStop}%
\bibitem [{\citenamefont {Shimizu}\ \emph {et~al.}(2010)\citenamefont
  {Shimizu}, \citenamefont {Tu},\ and\ \citenamefont {Berg}}]{shimizu10}%
  \BibitemOpen
  \bibfield  {author} {\bibinfo {author} {\bibfnamefont {T.~S.}\ \bibnamefont
  {Shimizu}}, \bibinfo {author} {\bibfnamefont {Y.}~\bibnamefont {Tu}}, \ and\
  \bibinfo {author} {\bibfnamefont {H.~C.}\ \bibnamefont {Berg}},\ }\href
  {\doibase 10.1038/msb.2010.37} {\bibfield  {journal} {\bibinfo  {journal}
  {Mol Syst Biol}\ }\textbf {\bibinfo {volume} {6}},\ \bibinfo {pages} {382}
  (\bibinfo {year} {2010})}\BibitemShut {NoStop}%
\bibitem [{\citenamefont {Vaknin}\ and\ \citenamefont {Berg}(2007)}]{vaknin07}%
  \BibitemOpen
  \bibfield  {author} {\bibinfo {author} {\bibfnamefont {A.}~\bibnamefont
  {Vaknin}}\ and\ \bibinfo {author} {\bibfnamefont {H.~C.}\ \bibnamefont
  {Berg}},\ }\href {\doibase 10.1016/j.jmb.2006.12.024} {\bibfield  {journal}
  {\bibinfo  {journal} {Journal of Molecular Biology}\ }\textbf {\bibinfo
  {volume} {366}},\ \bibinfo {pages} {1416} (\bibinfo {year}
  {2007})}\BibitemShut {NoStop}%
\bibitem [{\citenamefont {Li}\ and\ \citenamefont {Hazelbauer}(2004)}]{li04}%
  \BibitemOpen
  \bibfield  {author} {\bibinfo {author} {\bibfnamefont {M.}~\bibnamefont
  {Li}}\ and\ \bibinfo {author} {\bibfnamefont {G.~L.}\ \bibnamefont
  {Hazelbauer}},\ }\href {\doibase 10.1128/JB.186.12.3687-3694.2004} {\bibfield
   {journal} {\bibinfo  {journal} {Journal of Bacteriology}\ }\textbf {\bibinfo
  {volume} {186}},\ \bibinfo {pages} {3687} (\bibinfo {year}
  {2004})}\BibitemShut {NoStop}%
\bibitem [{\citenamefont {Skoge}\ \emph {et~al.}(2013)\citenamefont {Skoge},
  \citenamefont {Naqvi}, \citenamefont {Meir},\ and\ \citenamefont
  {Wingreen}}]{skoge13}%
  \BibitemOpen
  \bibfield  {author} {\bibinfo {author} {\bibfnamefont {M.}~\bibnamefont
  {Skoge}}, \bibinfo {author} {\bibfnamefont {S.}~\bibnamefont {Naqvi}},
  \bibinfo {author} {\bibfnamefont {Y.}~\bibnamefont {Meir}}, \ and\ \bibinfo
  {author} {\bibfnamefont {N.~S.}\ \bibnamefont {Wingreen}},\ }\href {\doibase
  10.1103/PhysRevLett.110.248102} {\bibfield  {journal} {\bibinfo  {journal}
  {Physical Review Letters}\ }\textbf {\bibinfo {volume} {110}},\ \bibinfo
  {pages} {248102} (\bibinfo {year} {2013})}\BibitemShut {NoStop}%
\bibitem [{\citenamefont {Wang}\ \emph {et~al.}(2007)\citenamefont {Wang},
  \citenamefont {Rappel}, \citenamefont {Kerr},\ and\ \citenamefont
  {Levine}}]{wang07}%
  \BibitemOpen
  \bibfield  {author} {\bibinfo {author} {\bibfnamefont {K.}~\bibnamefont
  {Wang}}, \bibinfo {author} {\bibfnamefont {W.-J.}\ \bibnamefont {Rappel}},
  \bibinfo {author} {\bibfnamefont {R.}~\bibnamefont {Kerr}}, \ and\ \bibinfo
  {author} {\bibfnamefont {H.}~\bibnamefont {Levine}},\ }\href@noop {}
  {\bibfield  {journal} {\bibinfo  {journal} {Phys Rev E Stat Nonlin Soft
  Matter Phys}\ }\textbf {\bibinfo {volume} {75}},\ \bibinfo {pages} {061905}
  (\bibinfo {year} {2007})}\BibitemShut {NoStop}%
\bibitem [{\citenamefont {Danielson}\ \emph {et~al.}(1994)\citenamefont
  {Danielson}, \citenamefont {Biemann}, \citenamefont {Koshland},\ and\
  \citenamefont {Falke}}]{danielson94}%
  \BibitemOpen
  \bibfield  {author} {\bibinfo {author} {\bibfnamefont {M.~A.}\ \bibnamefont
  {Danielson}}, \bibinfo {author} {\bibfnamefont {H.-P.}\ \bibnamefont
  {Biemann}}, \bibinfo {author} {\bibfnamefont {D.~E.}\ \bibnamefont
  {Koshland}}, \ and\ \bibinfo {author} {\bibfnamefont {J.~J.}\ \bibnamefont
  {Falke}},\ }\href {http://www.ncbi.nlm.nih.gov/pmc/articles/PMC2897698/}
  {\bibfield  {journal} {\bibinfo  {journal} {Biochemistry}\ }\textbf {\bibinfo
  {volume} {33}},\ \bibinfo {pages} {6100} (\bibinfo {year}
  {1994})}\BibitemShut {NoStop}%
\bibitem [{\citenamefont {van Albada}\ and\ \citenamefont {ten
  Wolde}(2009)}]{albada09a}%
  \BibitemOpen
  \bibfield  {author} {\bibinfo {author} {\bibfnamefont {S.~B.}\ \bibnamefont
  {van Albada}}\ and\ \bibinfo {author} {\bibfnamefont {P.~R.}\ \bibnamefont
  {ten Wolde}},\ }\href {\doibase 10.1371/journal.pcbi.1000378} {\bibfield
  {journal} {\bibinfo  {journal} {PLoS Comput Biol}\ }\textbf {\bibinfo
  {volume} {5}},\ \bibinfo {pages} {e1000378} (\bibinfo {year}
  {2009})}\BibitemShut {NoStop}%
\bibitem [{\citenamefont {Cheong}\ \emph {et~al.}(2011)\citenamefont {Cheong},
  \citenamefont {Rhee}, \citenamefont {Wang}, \citenamefont {Nemenman},\ and\
  \citenamefont {Levchenko}}]{cheong11a}%
  \BibitemOpen
  \bibfield  {author} {\bibinfo {author} {\bibfnamefont {R.}~\bibnamefont
  {Cheong}}, \bibinfo {author} {\bibfnamefont {A.}~\bibnamefont {Rhee}},
  \bibinfo {author} {\bibfnamefont {C.~J.}\ \bibnamefont {Wang}}, \bibinfo
  {author} {\bibfnamefont {I.}~\bibnamefont {Nemenman}}, \ and\ \bibinfo
  {author} {\bibfnamefont {A.}~\bibnamefont {Levchenko}},\ }\href {\doibase
  10.1126/science.1204553} {\bibfield  {journal} {\bibinfo  {journal}
  {Science}\ }\textbf {\bibinfo {volume} {334}},\ \bibinfo {pages} {354}
  (\bibinfo {year} {2011})}\BibitemShut {NoStop}%
\end{thebibliography}%


\clearpage
\appendix

\renewcommand{\theequation}{S\arabic{equation}}
\setcounter{equation}{0}

\renewcommand{\thefigure}{S\arabic{figure}}
\setcounter{figure}{0}

\setcounter{page}{1}
\renewcommand{\thepage}{S\arabic{page}}

\section{\LARGE {Supplementary Material}}

\subsection{Predictive information limits fitness when cells adapt with a
	delay}

The predictive mutual information between input and output is a biologically
relevant measure~\cite{taylor07,bialek12,bergstrom05} for the performance of
the system. Imagine a cell which instantaneously senses a slowly varying
nutrient concentration $s$, with intracellular output $x$, and as a result,
grows at a rate $g(x,s)$ that is maximal for some optimally adapted value of
$x(s)$. In this setting, it was shown~\cite{taylor07} that the instantaneous
mutual information $I[x,s]$ sets an upper limit for the achievable growth rate.

Now imagine that the cell's adaptive response (\emph{e.g}, enzyme production)
requires finite time $\tau$ to mount. Since only available enzymes can process
nutrients, the growth rate lags behind the output $x$, as $g(t+\tau) = g[x(t),
s(t+\tau)]$. Repeating the argument of~\cite{taylor07}, the predictive
information $I[x,s_\tau]=I[x(t),s(t+\tau)]$ now limits the achievable growth
rate. This makes maximizing $I[x,s_\tau]$ a plausible design goal for dynamic
sensory networks.

We have considered a fixed delay time $\tau$ for simplicity, leading to the
mutual information $I[x,s_\tau]$ as a relevant information measure, which is
distinct from both the predictive information between entire past and future
trajectories \cite{bialek01} and the information rate between input and output
trajectories \cite{tostevin09}. Distributed delays for the adaptive cellular
response of the cell would correspond to yet different predictive information
definitions in which future time points are weighted according to a delay
distribution; such refinements are left for future work.

\subsection{Instantaneous response is optimal for noiseless Markovian signals}

Consider first a general input signal, Markovian or not. Denote the full signal
history up to but excluding $t$ by $\hsl$, the present signal by $s=s(t)$, the
present output by $x=x(t)$, and a future signal by $s_\tau = s(t+\tau)$,
respectively.  We require that the output does not feed back onto the signal.
This implies that future signal values are independent of the response for
given signal history: $p(s_\tau|x, \hsl, s, t) = p(s_\tau|\hsl, s, t)$.

We can then expand the predictive two-point distribution
$p(s_\tau,x|s, t)$ over the driving history:
\begin{flalign}\label{eq:jointcond1}
&p(s_\tau,x|s, t)
= \int \mathcal D\hsl p(s_\tau, x, \hsl|s, t)
	\nonumber\\
&= \int \mathcal D\hsl p(s_\tau| x, \hsl,s, t)
p(x|\hsl, s, t)p(\hsl|s, t)
	\nonumber\\
&= \int \mathcal D\hsl p(s_\tau| \hsl,s, t)
p(x|\hsl, s, t)p(\hsl|s, t),
\end{flalign}
where the second equality uses the basic rule $p(x,y|z)=p(x|y,z)p(y|z)$, and the
last equality uses the absence of feedback.

If the response is instantaneous, we have the relation $p(x|\hsl, s, t) =
p(x|s,t)$; if the input is Markovian, $p(s_\tau|\hsl, s,t) =
p(s_\tau|s,t)$. In either case we can integrate
Eq.~\ref{eq:jointcond1} over $\hsl$ and obtain
\begin{equation}\label{eq:jointcond2}
p(s_\tau,x|s, t) =  p(x|s, t)p(s_\tau|s, t).
\end{equation}
So, if the input $s$ is Markovian, history-dependent and instantaneous
responders behave the same: $s_\tau$ and $x$ are independent when conditioned
on $s$. In other words, the variables $x \leftrightarrow s \leftrightarrow
s_\tau$ form a Markov chain~\cite{cover12}. They therefore obey the data
processing inequality $I[x,s_\tau]\leq I[s,s_\tau]$.

This implies that predictions based on measuring $x$ cannot surpass those based
on the current input $s$, even if $x$ depends on the history $\hsl$ in
arbitrarily complicated ways. Thus, an instantaneous responder for which $x$
faithfully tracks the current input $s$, can already achieve the maximal
performance $I[s,s_\tau]$. Note that this requires $s$ to be noiseless; in
contrast, degraded Markovian signals can be better predicted with memory,
as discussed in the main text.

\subsection{Optimal prediction of Gaussian signals \\ by linear networks}

In this section we consider optimal linear prediction strategies for input
concentration signals that are stationary Gaussian processes. An input signal
$s(t)$ is presented to the responder in form of a ligand number $\ell(t)$
subject to molecular noise. Ligands are sensed by a sensory network which
operates in a linear regime but is otherwise arbitrary, possibly including
multiple stages and feedback loops. We impose the restriction that the
network does not have any influence on the input, for instance, by sequestering
ligands and thereby reducing the current available ligand number. Since we are
interested in how the accuracy of prediction depends on the noise and the
correlations in the input signal, we focus on responders in the
deterministic limit.

In this linear regime, we can expand the input $s(t)$ and degraded input
$\ell(t)$ around their steady state values, and subtract the latter, so that in
the following $\E{s}=0$ and $\E{\ell}=0$ without restriction. The degraded
input is modeled as
\begin{equation}\label{eq:whitelnoise}
	\ell(t) = s+\xi,
\end{equation}
where $\xi$ is independent, Gaussian white noise with $\E{\xi(t)\xi(t')} =
\sigma_s^2\vartheta^2 \delta(t-t')$ and $\sigma_s$ denotes the standard
deviation of the pure, non-noisy input process. The white noise $\xi$
approximates a physical noise process $n$ with variance $\sigma_n^2$ and finite
but short correlation time $\tau_n$. Its effect on the final output of the
signaling network is determined by the integrated strength $\int_{0}^{\infty}
\E{n(0)n(t)}dt = \sigma_n^2\tau_n$. 
The integrated strength of $\xi$ is $\sigma_s^2\vartheta^2$. Thus,
$\vartheta^2$ measures the integrated noise strength relative to the signal
variance, and has units of time. 

Since we disregard responder noise, the absolute amplitude of the signal
$\sigma_s$ is irrelevant, as will be clarified shortly. The sensory output is
then given as a convolution
\begin{equation}
    x(t) = \int_{-\infty}^{t} \ell(t'') k(t-t'') dt''
		= \int_{0}^{\infty} \ell(t-t'') k(t'') dt'',
\end{equation}
where memory of past signal values is encoded by the
linear response kernel $k$.
The output $x$ is correlated with the input. Here and in the following, we use
$c_{xy}=\E{xy}$ and $r_{xy} = c_{xy}/\sigma_x\sigma_y$ to denote covariance
functions and correlation functions, respectively, and abbreviate
$r_{xx}=r_x,c_{xx}=c_x$. Then the predictive input-output correlation
is given by
\begin{align}
	r_{x s} (\tau) &= r_{xs_\tau} = \frac{c_{xs_\tau}}{\sigma_x\sigma_s}
		\nonumber\\
		&= \frac{\int_{0}^{\infty}
			\E{[s(t'-t'')+\xi(t'-t'')]k(t'')s(t'+\tau)} dt''}{\sigma_x\sigma_s}
		\nonumber\\
		&= \frac{\int_{0}^{\infty}r_{s}(t'+\tau)k(t') dt'}{\sigma_x/\sigma_s}
		\nonumber\\
		&= \frac{\int_{0}^{\infty}r_{s}(t'+\tau)k(t') dt'}
		{\bigl[\int_{0}^{\infty}
			k(t)[r_s(t-t')+\vartheta^2\delta(t-t')]k(t')dtdt'
                        \bigr]^{1/2}}.\nonumber\\
                      &\equiv \frac{\Psi}{[\Sigma + \Xi]^{1/2}}
		\label{eq:iocorrsupp}
\end{align}
Importantly, since $s$ and $x$ are jointly Gaussian, the predictive information
evaluates to $I[x,s_\tau]=-\log[1-r_{x s}(\tau)^2]/2$, a simple function of the
cross-correlation, see also~\cite{tostevin09,tostevin10}. Since $r_{xs}$ does
not depend on the absolute signal amplitude $\sigma_s$, neither does the
predictive information.
The numerator of the last equation defines the
overlap integral between the current output $x(t)$ and
the future input $s(t+\tau)$, normalized by the input variance
$\sigma^2_s$: $\Psi \equiv c_{x s_\tau} / \sigma^2_s=
\int_{0}^{\infty}r_{s}(t'+\tau)k(t') dt'$. The denominator defines
the part of the normalized output variance $\sigma^2_x/\sigma^2_s$
that is due to the past input signal $s(t)$, $\Sigma\equiv \int_0^\infty
k(t)r_s(t-t')k(t')dtdt'$, and the part which is due to past noise $\xi (t)$, $\Xi\equiv \vartheta^2 \int_0^\infty
k(t)^2dt$.

\subsection{Exponential response kernels}
In this section, we consider systems with exponential response
kernels. The input signals are either Markovian or non-Markovian, with
the latter being generated from the harmonic-oscillator model as described in
the main text.

\subsubsection{Optimal response speeds}

We can readily evaluate $r_{x s}(\tau)$ for systems with
exponential response kernels, $k\propto
\exp(-\mu t)$. For Markovian (M) Ornstein-Uhlenbeck signals with $r_s(t) =
\exp(-\lambda t)$ we obtain
\begin{equation}\label{eq:rnaOU}
	r_{x s}^\text{M}(\tau) = \exp(-\lambda \tau)
	\bigl[(\mu+\lambda)(1+\vartheta^2 (\mu+\lambda)/2)/\mu\bigr]^{-1/2}.
\end{equation}
Maximizing the correlation yields the optimal response speed
$\mu_\text{ti}=(2\lambda/\vartheta^2+\lambda^2)^{1/2}$.

For a NM signal with damping coefficient $\eta$, the correlation
coefficient, after some algebra, results as 
\begin{gather}\label{eq:rnaHO}
%
r_{x s}^\text{NM}(\tau) = \hspace*{14em} \\
\frac{e^{-\frac{1}{2} \tau \eta_+}[2-2\eta_{-}(\eta+\mu)] -
	e^{-\frac{1}{2} \tau \eta_-}[2-2\eta_{+}(\eta+\mu)]}
{(\eta\mu+\mu^2+1)\bar\eta
	\bigl[\frac{8}{a_+} + \frac{8}{a_-}+ \frac{-2\eta^2 \vartheta^2 \mu^2+4 \eta
		+2\vartheta^2 \left(\mu^2+1\right)^2}{-\left(\eta^2-2\right)
		\mu^3+\mu^5+\mu }	\bigr]^{1/2}},
\nonumber
\end{gather}
where
\begin{align}
a_\pm &= \sqrt{\eta \eta_{\pm}-1} [\eta ^2\eta_\pm+2 (\mu ^2 + 1) \bar\eta -
4\eta ]\text{, }\nonumber\\
\eta_\pm &= \tfrac12 (\eta\pm\bar\eta)\text{, and } \bar\eta=\sqrt{\eta^2-4}
\nonumber.
\end{align}
Maximization of $(r_{x s}^\text{NM})^2$ with respect to the response speed $\mu$
can be done numerically and yields a characteristic discontinuity at finite
$\tau$; see Fig.~\ref{fig:optkernels}B of the main text.

\begin{figure}[tb]
	\begin{center}
		\includegraphics[width=.6\columnwidth]{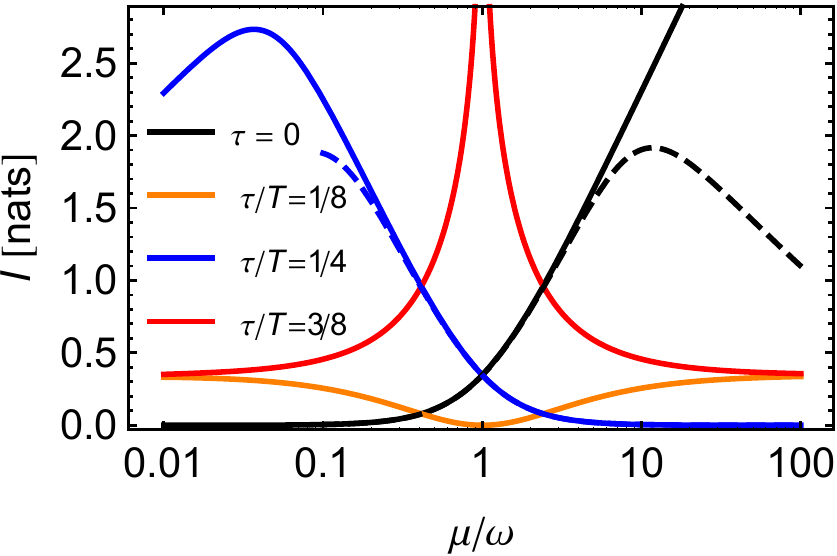}
		\caption{\label{fig:lineardetlim}%
			Predictive information of exponential responders for highly
			correlated NM signals with $\eta=10^{-4}$. Noise strength
			$\vartheta=0,0.05$, solid and dashed lines, respectively. Cf.~also
			Fig.~\ref{fig:optkernels}D}
	\end{center}
\end{figure}

The resulting predictive informations $I$ are plotted in
Fig.~\ref{fig:linearintro} of the main text for signals with moderate intrinsic
stochasticity $\eta>0.5$. For comparison, Fig.~\ref{fig:lineardetlim} shows the
predictive information achieved for long-range correlated non-Markovian (NM)
harmonic oscillator signal at $\eta=10^{-4}$. Without noise and for prediction
interval $\tau>0.2$, a slow response can induce a delay $\Lambda$ that leads to
optimal anticorrelation with the future signal, by achieving anti-phase
matching $\Lambda+\tau=T/2$ (peaks in Fig.~\ref{fig:lineardetlim}). In fact,
when taking $\eta\to0$, the mutual information between these two perfectly
anticorrelated time points diverges. Input noise (dashed lines) regularizes the
divergence, leaving pronounced maxima at the delays that produce the best
(anti)phase matching.

\subsubsection{Scaling argument: Time integration leads to an optimal response
	speed $\mu_\text{ti}$ for noisy signals}

We consider the scaling of Eq.~\ref{eq:iocorr} (or Eq.~\ref{eq:iocorrsupp})
with the effective integration time $t_\text{int}$ in a simplified setting. Let
$r_s$ and $k$ be positive, monotonically decreasing and of finite support.
Specifically, let $k(0)=1$, and $k(t)=0$ for $t>t_\text{int}$, the integration
time, and let $r_s(t)=0$ for $|t|>t_s$, the correlation time.

Then, in the regime $t_\text{int}<t_s-\tau$, the overlap
$\Psi(\tau)=\int_0^\infty r_s(t+\tau)k(t)dt$ in the numerator of
Eq.~\ref{eq:iocorr} increases roughly linearly with increasing $t_\text{int}$.
In the denominator, the contribution of past signals to the output variance
$\Sigma = \int k r_s k \propto t_\text{int}^2$  while the noise contribution
$\Xi = \vartheta^2\int k^2\propto t_\text{int}$. Thus for short integration
times, the noise contribution dominates and $r_{xs_\tau}^2$ increases $\propto
t_\text{int}/\vartheta^2$ due to noise averaging. This initial scaling regime
exists only for noisy signals $\vartheta>0$; it is the reason why short-time
integration benefits predictions.

When the integration time exceeds the signal correlation time, a new regime
appears where the overlap integral $\Psi$  saturates to a constant, but both
$\Sigma\propto t_\text{int}$ and $\Xi\propto t_\text{int}$ continue to grow, so
that $r_{xs_\tau}^2\propto t_\text{int}^{-1}$. This decrease in predictive
power at long integration times reflects the fact that very long signal
integration confounds predictions by including uncorrelated past signals; this
scaling regime therefore exists with or without noise. In effect, uncorrelated,
past signal values are a source of noise, which hampers prediction.

We conclude that in the presence of noise, there exists an optimal response
speed $ 1/t_\text{int}^\vartheta < \mu_\text{ti} < 1/t_s$ that allows noise
averaging but excludes confounding past signals.

In the above analysis, we have modeled the noise as white Gaussian noise. The
same arguments still apply when the correlation time of the noise is finite,
but shorter than the signal correlation time. In that case, there also exists
an optimal response time that arises from the trade-off between noise averaging
and excluding past signal values uncorrelated with the future. The situation is
different in the opposite limiting case where the noise correlation time is
longer than that of the signal. Here we expect that an adaptive filter may be
beneficial for predicting the future; we leave this for future work.

\subsubsection{Exploitation of non-monotonic signal autocorrelations yields an
	optimal response speed $\mu_*$}

Non-exponential autocorrelations can generate an optimal response speed $\mu_*$
even without noise. Without noise, $t_\text{int}^\vartheta=0$, and the
arguments in the previous section do not allow the conclusion that there is an
upper bound for the optimal response speed. In fact, we know already that when
the input signal is Markovian, an instantaneous response is optimal.

However, when the input signal is non-Markovian, there exists a finite response
speed even in the absence of noise in the input.  One way to see this is to
consider only  exponential responders $k(t)=\exp(-\mu t)$ and expand for high
speeds $\mu$. One obtains
\begin{equation}
	r_{xs_\tau}^2 =
	r_s(\tau)^2\bigl[1 +
	\tfrac{2}{\mu}\bigl(\tfrac{r_s'(\tau)}{r_s(\tau)}-\tfrac{r_s'(0)}{r_s(0)}\bigr)
	\bigr] +O(\mu^{-2}).
\end{equation}
One sees that in the case of exponential $r_s$, the sub-leading $1/\mu$-term
vanishes exactly; this is compatible with the fact that the optimal
response for Markovian noiseless signals is instantaneous, as discussed above.
In contrast, when the autocorrelation function $r_s (t)$ of the input signal is
non-monotonic or becomes negative, lagging responders with finite $\mu$ can
make the $1/\mu$ term positive and thus improve predictions over the
instantaneous response $\mu\to\infty$. This shows that instantaneous responders
are not globally optimal predictors at least for some non-Markovian noiseless
signals.

\begin{figure}[tb]
	\begin{center}
		\includegraphics[width=.7\columnwidth]{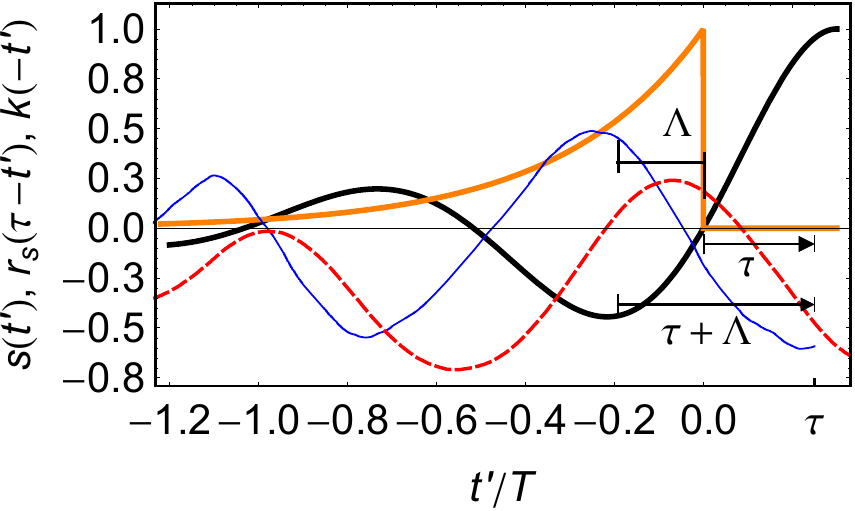}
		\caption{\label{fig:kernelcartoon}%
			Exploiting signal correlations to predict the future. For the linear
			networks studied here, the output $x(t)$ (dashed, red) at time $t = 0$ is
			given by $x(0) = \int_{-\infty}^0 k(- t^\prime) s(t^\prime)$. Hence, the
			kernel $k(-t')$ (orange) integrates the signal $s(t^\prime)$ (blue) over
			past time points $t^\prime<0$. The challenge is to pick up the signals
			$s(t^\prime)$ from the past that are most correlated with the future signal
			$s(\tau)$ that is to be predicted. An instantaneous kernel would include
			signals around $t^\prime = 0$ which contain little information about
			$s(\tau)$, as can be seen from the zero-crossing of the signal
			autocorrelation function $r_s(\tau-t^\prime)$ (black) at $t^\prime = 0$. In
			contrast, the lagging kernel $k=k_*$, shown in orange, emphasizes the
			negative lobe of $r_s$, picking up signal values from the past that are
			strongly anticorrelated with the future signal $s(\tau)$.  This maximizes
			predictive power. Effectively, this kernel generates a response $x(t)$ that
			lags behind the input $s(t)$ with a lag $\Lambda$, so that the current
			output $x(0)$ reflects the past input $s(-\Lambda)$ rather than the current
			input $s(0)$. The optimal lag obeys $\tau + \Lambda \simeq T/2$, with $T$
			the period of the signal oscillation: $r_s(\Lambda + \tau)$ is, indeed,
			strongly negative. In this example, $\eta=0.5$, so that the signal $s(t)$
			is oscillatory.}
	\end{center}
\end{figure}

Fig.~\ref{fig:kernelcartoon} gives an intuitive explanation of the mechanism of
exploitation of correlated past signals. In essence, a lagging kernel weights
heavily the past signal values $s(t')$ at those time points that are strongly
correlated with the prediction target $s(t+\tau)$, as reported by the shifted
autocorrelation $r_s(t+\tau)$.

In particular, when $r_s(t)$ is non-monotonic, the overlap $\Psi(\tau)$ and the
signal-induced variance $\Sigma$ depend strongly on the shape of the kernel and
autocorrelation functions. $r_{xs_\tau}^2$, and $I$, is maximized when the
kernel overlaps well with the back-shifted autocorrelation, increasing
$\Psi(\tau)^2$, while overlapping weakly with forward-shifted autocorrelations,
decreasing $\Sigma = \int \Psi(-t')k(t')dt'$. Thus, an optimal kernel selects
past signals that are correlated with the future signal while rejecting
confounding past signals. This basic strategy continues to be effective at
finite noise levels, where noise averaging and exploitation of correlations are
combined via Eq.~\ref{eq:iocorr}, as seen in Fig.~\ref{fig:linearintro}D.

\subsection{Wiener-Kolmogorov filtering theory for predicting biochemical
	signals}

We are interested in finding a response kernel with optimal predictive power.
In signal processing, statistically optimal filters are well known; in
particular, the Wiener filter~\cite{kolmogorov92,wiener50} is the linear
response kernel $k_*$ that minimizes the mean squared error between a general
output $o$ and input $i$, $e_{io}=\E{(i-o)^2}=\min$. This criterion is
different from optimal predictive power, $I[i,o]=\max$. However, in the case of
Gaussian signals, both are closely related.

\paragraph{For Gaussian signals, the Wiener filter is optimally informative}
In order to minimize the mean squared error
\begin{equation}
	e_{io} = \sigma_i^2 + \sigma_o^2 - 2 \sigma_i \sigma_o r_{io}.
\end{equation}
for given input statistics, we are free to start by rescaling the output
variable (by adjusting the kernel amplitude). The optimal scale is attained
when $d e_{io}/d{\sigma_o} = 0$, or $\sigma_o = \sigma_i r_{io}$ (recall that
$r_{io}$ does not depend on the output scale). The latter equation makes sense
only if $r_{io} > 0$; we can ensure this by inverting the kernel if necessary.
Inserting this result, $e_{io} = \sigma_i^2 (1-r_{io}^2)$. Since the input
amplitude is fixed, we conclude that $e_{io}=\min \Rightarrow r_{io}^2 = \max$.
Conversely, for Gaussian signals, $I=\max \Leftrightarrow r_{io}^2=\max$.
Together, this shows that the predictive information is maximized by any kernel
in the family $\{\alpha k_*\}_{\alpha \neq 0}$, which maximally correlate
($\alpha > 0$) or anticorrelate ($\alpha < 0$) with the input. We can thus
exploit the Wiener-Kolmogorov filter theory to construct optimally informative
response functions for Gaussian signals.

\paragraph{Construction of the Wiener-Kolmogorov causal filter}
The causal Wiener filter is the optimally predictive kernel $k_*$ which depends
only on past signal values. We briefly recall the construction of $k_*$ for
given signal correlations, noise levels and prediction interval. Let the
Fourier transform of the degraded input autocovariance, $ c_\ell(\omega) \equiv
\mathcal{F}[c_\ell](\omega) = \int_{-\infty}^\infty c_\ell(t) e^{-i\omega
	t}dt$. Define the causal part of a function $f(\omega)$ as $[f]_+(\omega) =
\mathcal{F}[\theta \mathcal{F}^{-1}[f]](\omega)$ where $\theta$ is
multiplication with the unit step function. Now, find a Wiener-Hopf
factorization $c_\ell(\omega)=m_+m_-$, which is defined by the properties
$m_-(\omega)=m_+(\omega)^*=m_+(-\omega)$ which ensure that $m_\pm$ are real
functions in the time domain; and by the requirement that $m_+$ and $1/m_+$ be
causal, that is, their inverse transforms vanish for all $t<0$, which can be
written as $m_+=[m_+]_+$ and $1/m_+=[1/m_+]_+$, respectively. Then the causal
Wiener filter is given by~\cite{wiener50}
\begin{equation}
	k_*(\omega) = \frac{1}{m_+(\omega)}
	\Bigl[\frac{c_{\ell s_\tau}(\omega)}{m_-(\omega)}\Bigr]_+.
\end{equation}
This general result simplifies in our case of independent white noise,
Eq.~\ref{eq:whitelnoise}. Here,
$c_\ell(t) = \sigma_s^2 [r_s(t)+\vartheta^2\delta(t)]$, so that $c_{\ell
	s}(\omega) = c_s(\omega) = c_\ell(\omega) - \sigma_s^2\vartheta^2$. Noting
that the
predictive cross-covariance satisfies $c_{\ell s_\tau}(\omega) = c_{\ell
	s}(\omega) e^{i\omega\tau}$, we obtain
\begin{equation}\label{eq:wienersimp}
	k_* = \frac{1}{m_+}
	\Bigl[\Bigl(m_+ - \frac{\sigma_s^2\vartheta^2}{m_-}\Bigr)e^{i\omega\tau}\Bigr]_+
	= m_+^{-1}\bigl[m_+ e^{i\omega\tau}\bigr]_+,
\end{equation}
where we substituted the factorization of $c_\ell$ and in the last equality, used
the fact that $1/m_-$ and even more so, the back-shifted $e^{i\omega\tau}/m_-$,
has no causal part.

\paragraph{When is the instantaneous response optimally predictive?}
With this general expression we can answer the question under what conditions
the instantaneous response is optimal. According to Eq.~\ref{eq:wienersimp},
this happens exactly when $k_*(\omega)=m_+^{-1}\bigl[m_+
e^{i\omega\tau}\bigr]_+=a$ is a real constant, since then
$k_*(t)\propto\delta(t)$. This occurs if and only if $m_+(t+\tau)/m_+(t) = a$
for all $t\geq0$. By writing $c_\ell(t)=\int m_+(t')m_-(t-t')dt'$, this can be
seen to imply
\begin{equation}\label{eq:deltacond}
c_\ell(t+\tau) = a c_\ell(t), \,t \geq 0,
\end{equation}
where a bounded signal requires $|a|\leq 1$. We conclude that the
input correlation $c_\ell(t)$ must decay exponentially, possibly modulated by
oscillations; in the latter case, the forecast interval $\tau$ must be a
multiple of the period. 

This condition is quite restrictive. First, Eq.~\ref{eq:deltacond} pertains to
the degraded input $c_\ell$, not the signal $c_s$ itself. This means that an
instantaneous responder cannot be optimal when there is noise in the input
signal, since if $\vartheta > 0$, $c_\ell(t)$ would exhibit a $\delta$-peak at
$t=0$, violating Eq.~\ref{eq:deltacond}. This observation agrees also with the
scaling analysis above which found that for noisy input, short-time integration
is beneficial. Second, if the input decays with exponentially damped
oscillations, then Eq.~\ref{eq:deltacond} requires that the prediction interval
is a multiple of the input period. This requirement corresponds to predicting
only exactly phase-related future time points via a simple phase matching
mechanism.

We conclude that an instantaneous, $\delta$-shaped responder is optimal only in
the contrived limit in which (a) the input is noiseless; and (b) the input
decays exponentially (the Markovian signal), or it decays with exponentially
damped oscillations and the forecast interval $\tau$ is a multiple of the input
period $T$.

\subsection{Computation of optimal kernels}

We now give explicit expressions for optimally predictive kernels
for several input signal types.

\subsubsection{Optimal kernels for Markovian signals}

The Markovian (M) signal autocorrelation reads $r_s(t)=\exp(-\lambda |t|)$ for
some positive damping coefficient $\lambda$. The power spectral density of the
degraded input is $c_\ell(\omega) = \frac{2\lambda}{\lambda^2+\omega^2} +
\vartheta^2$, for which we find a Wiener-Hopf factorization
\begin{equation} m_\pm(\omega) =
	\frac{\zeta\pm i\vartheta\omega}{\lambda\pm i\omega};\;
	\zeta=\sqrt{2\lambda+\lambda^2\vartheta^2}
\end{equation}
and finally the result
\begin{equation}
	k_*^\text{M}(\omega)= e^{-\lambda\tau}
	\frac{\zeta-\lambda\vartheta} {\zeta+i\vartheta\omega}.
\end{equation}
In the noiseless limit $\vartheta\to0$, the kernel reduces to a constant
$e^{-\lambda\tau}$ in the frequency domain, confirming that indeed for
noiseless Markovian signals, an instantaneous responder has optimal predictive
power. (Note that this is compatible with Eq.~\ref{eq:deltacond}.) For positive
noise strength $\vartheta>0$, the optimal kernel in the time domain reads
\begin{equation}
	k_*^\text{M}(t)\propto \theta(t) e^{-\mu_\text{ti} t}\text{, where }
	\mu_\text{ti}=\sqrt{\lambda^2+2\lambda/\vartheta^2}.
\end{equation}
That is, the input noise is filtered optimally by an exponential moving
average; the averaging time constant $1/\mu_\text{ti}$ arises from time
integration, since the Markovian signal does not provide extra information in
past signal values. The time constant increases from 0 for weak noise towards
$1/\lambda$, the input signal correlation time, for strong noise.
$\mu_\text{ti}$ (necessarily) agrees with the optimal response speed found
previously, when optimization was restricted to exponential responders only.
Thus, a lagging response improves prediction of noisy Markovian signals, by
allowing a better estimate of the current signal value $s(t)$ through
averaging; this is the best strategy possible with a linear responder.

\subsubsection{Optimal kernels for noiseless non-Markovian signals}

In the NM case, defined as in the main text by the Langevin equations
$\partial_{\omega t} q = p,\; \partial_{\omega t}p = -q -\eta p + \sqrt{2\eta}
\psi$, we set $q \equiv s$. \r{Note that the process $(q(t),p(t))$ is indeed a
	Gaussian process; this follows from the fact that future signals are
	generated as a linear combination of present signals and Gaussian noise, and
	that Gaussian random variables are stable under linear combination. Since the
	Langevin equations do not depend on time explicitly, after some transient
	period, the process is stationary.}

The power spectral density of the degraded input becomes
\begin{equation}\label{eq:rlhonoise}
	c_\ell(\omega) = \frac{2\eta}{\omega^4+(\eta^2-2)\omega^2+1} + \vartheta^2.
\end{equation}
Considering first the case without input noise $\vartheta=0$, we exploit the
fact that $m_+$ is proportional to the dynamic susceptibility of the harmonic
oscillator, to obtain $m_+(\omega) \propto (\omega^2 -  i \eta\omega -
1)^{-1}$, or in the time domain, $m_+(t)\propto \theta(t) e^{-\eta
	t/2}\sin(\eta' t/2)/\eta'$, where $\eta'=\sqrt{4-\eta^2}$. One then obtains
the optimal response
\begin{multline}\label{eq:khononoise}
	k_*^\text{NM} \propto
	\bigl( \cos\tfrac{\eta'\tau}2+ \tfrac{\eta}{\eta'}
	\sin\tfrac{\eta'\tau}2 \bigr)\delta(t)
	+ \bigl(\tfrac2{\eta'} \sin\tfrac{\eta'\tau}2\bigr)\dot\delta(t).
\end{multline}
Recalling that by definition,
\[\int
[a(\tau)\delta(t)-b(\tau)\dot\delta(t)]\ell(t-t')dt' =
a(\tau)\ell(t)+b(\tau)\dot\ell(t),\]
one sees that this response kernel implements prediction by linear
extrapolation. The network output is a linear combination of the current signal
value $\ell(t)$ and derivative $\dot \ell(t)$ whose coefficients $a(\tau)$ and
$b(\tau)$ depend on the signal statistics and prediction interval $\tau$. This
singular kernel can be thought of as the limit of a kernel with a  sharp
positive peak at $t\to 0$ immediately followed by a single sharp underswing;
such a kernel yields an output which effectively is a weighted sum of the
current input signal value and its derivative, the weights of which are
determined by the positive and negative lobe of the kernel. Although the
noiseless case is an oversimplified singular limit, the low-noise regime can be
understood in similar terms, see below and the low-noise kernel in
Fig.~\ref{fig:optkernelshighlow}A.

As a special case, notice that in the underdamped regime $\eta<2$, $\eta'$ is
real. When the prediction interval $\tau$ is chosen as a multiple of the signal
period $2\pi/\eta'$, the $\delta'$ term in Eq.~\ref{eq:khononoise} vanishes.
Then, the optimal response strategy reduces to instantaneous readout of the
signal. This finding reproduces the previous general result
Eq.~\ref{eq:deltacond}.

\subsubsection{Optimal kernels for non-Markovian signals with noise}
In the general NM case including input noise $\vartheta>0$, we may expect that
as for the Markovian signal, noise suppression through averaging becomes
relevant. We write Eq.~\ref{eq:rlhonoise} as a single fraction and factorize
the numerator into causal and anticausal parts. Combined with the noiseless
result, this yields
\begin{equation}
	m_+ \propto \frac{\vartheta\omega^2-i\zeta_2^+\omega-\zeta_1}
	{\omega^2 -  i \eta\omega - 1},
\end{equation}
where we define $\zeta_1=(2\eta+\vartheta^2)^{1/2}$ and $\zeta_2^\pm =
(\vartheta[(\eta^2-2)\vartheta \pm 2\zeta_1])^{1/2}$. With some algebra, the
optimal kernel in the Fourier domain can be written as
\begin{multline}\label{eq:khonoise}
	k_*^{\text{NM}} \propto
	\bigl[\vartheta \omega^2 - i \zeta_2^+ \omega - \zeta_1 \bigr]^{-1} \times \\
	\Bigl\{
		-i \omega  (\eta  -\zeta_2^+ / \vartheta ) \bigl[\tfrac{\zeta_2^+}{\eta'} \sin
   \tfrac{\tau  \eta'}{2} + \vartheta  \cos \tfrac{\tau  \eta '}{2}\bigr] + \\
	 + \tfrac{\eta \left(\vartheta +\sigma _1\right) - 2 \zeta_2^+}{\eta'}
	 \sin \tfrac{\tau  \eta'}{2} + (\zeta_1-\vartheta) \cos \tfrac{\tau  \eta'}{2}
\Bigr\}.
\end{multline}
The corresponding time-domain kernel is a superposition of exponential terms
with prefactors depending on signal, noise and the prediction interval; and
`rate constants' $\mu_*^\pm = (\zeta_2^+\pm\zeta_2^-)/(2\vartheta)$
determined by the poles in the complex $\omega$-plane of Eq.~\ref{eq:khonoise},
independent of the prediction interval $\tau$.

\begin{figure}[tb]
	\begin{center}
		\includegraphics[width=1.\columnwidth]{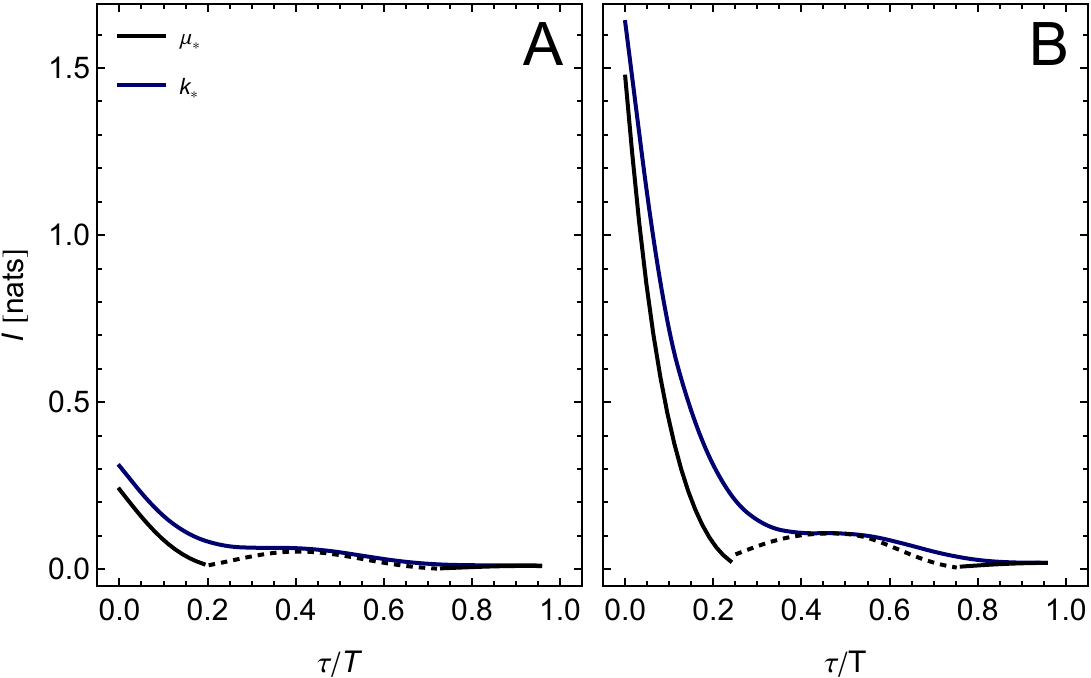}
		\caption{\label{fig:Ivstaunoise}%
			(A) Predictive information $I[x,s_\tau]$ for NM signals at
			$\eta=0.5,\vartheta=1$ for optimal exponential kernels $\mu^*=\mu^*(\tau)$
			(black; dashed region indicates $xs_\tau$ anticorrelation as in
			Fig.~\ref{fig:optkernels}B); and for the respective
			globally optimal kernel $k_*$, which is non-monotonic (top blue line). (B)
			as (A) but a low noise $\vartheta=0.1$. Note that (A) is the same as
			Fig.~\ref{fig:optkernelshighlow}B in the main text.}
	\end{center}
\end{figure}

The optimal exponential kernel with $\mu=\mu_*(\tau)$ improves performance as a
function of $\tau$ over any fixed exponential kernel; however, it still suffers
from the poor correspondence of signal and kernel shape when $\tau\simeq
\tau_c$,  where the correlation function is close to a zero-crossing and the
exponential response switches between correlated and anticorrelated output,
Fig.~\ref{fig:Ivstaunoise}A. The optimal, non-monotonic, kernel $k_*$ improves
predictions strongly over even the best-adapted exponential kernel,
Fig.~\ref{fig:Ivstaunoise}A, in particular around the critical prediction
interval $\tau_c$, where an oscillatory kernel can exploit signals on both
sides of the zero-crossing of $r_s$, avoiding precision loss due to
cancellations. At low noise, performance is better in general,
Fig.~\ref{fig:Ivstaunoise}B, but the advantage of the optimal kernel over
exponential kernels persists.

\subsubsection{General expression for optimal kernels at high noise} In the
Markovian case, we have seen that as the noise strength $\vartheta$ diverges,
the optimal kernel becomes proportional to the signal autocorrelation function.
This is in fact a general result which applies to arbitrary Gaussian input
signals. We recall that for white noise,
$c_\ell(\omega)=c_s(\omega)+\vartheta^2\sigma_s^2$, and expand the factors
$m_\pm$ as $m_\pm=\vartheta m_\pm^{0}
+\vartheta^{-1}m_\pm^{1}+O(\vartheta^{-3})$, so that comparing $c_\ell=m_+m_-$
order by order gives $\sigma_s^2=m_+^{0}m_-^{0}$ and
$c_s=m_+^{0}m_-^{1}+m_-^{0}m_+^{1}$. Since $\sigma_s$ is a real constant, this
implies $m_\pm^{0}=\sigma_s$, and $m_+^{1}=c_s/\sigma_s - m_-^{1}$.

Now recall that the kernel is given by Eq.~\ref{eq:wienersimp},
\begin{equation}
	k_*(\omega) = \frac
	{\bigl[(\sigma_s+\vartheta^{-2}m_+^{1})e^{i\omega\tau}\bigr]_+}
	{\sigma_s+\vartheta^{-2}m_+^{1}}
	= \frac
	{\vartheta^{-2}\bigl[(c_s/\sigma_s) e^{i\omega\tau}\bigr]_+}
	{\sigma_s+\vartheta^{-2}m_+^{1}},
\end{equation}
where the second equality follows after substituting $m_+^{1}$ and eliminating
all anticausal terms in the causal bracket. The last expression, to leading
order gives just $\vartheta^2 k_*=[c_s e^{i\omega\tau}]_+$.
This proves that the optimal kernel, to leading order for high noise, is
proportional to the signal autocorrelation function, back-shifted by the
prediction interval $\tau$, or $k_*(t)\propto \theta(t)c_s(t+\tau)$. We
note that the same result can also be obtained by functionally maximizing the
predictive correlation $r_{xs_\tau}$ with respect to the kernel in the high
noise limit.

\subsection{Networks that implement optimal kernels for oscillatory signals}

\begin{figure}[tb] \begin{center}
		\includegraphics[width=\columnwidth]{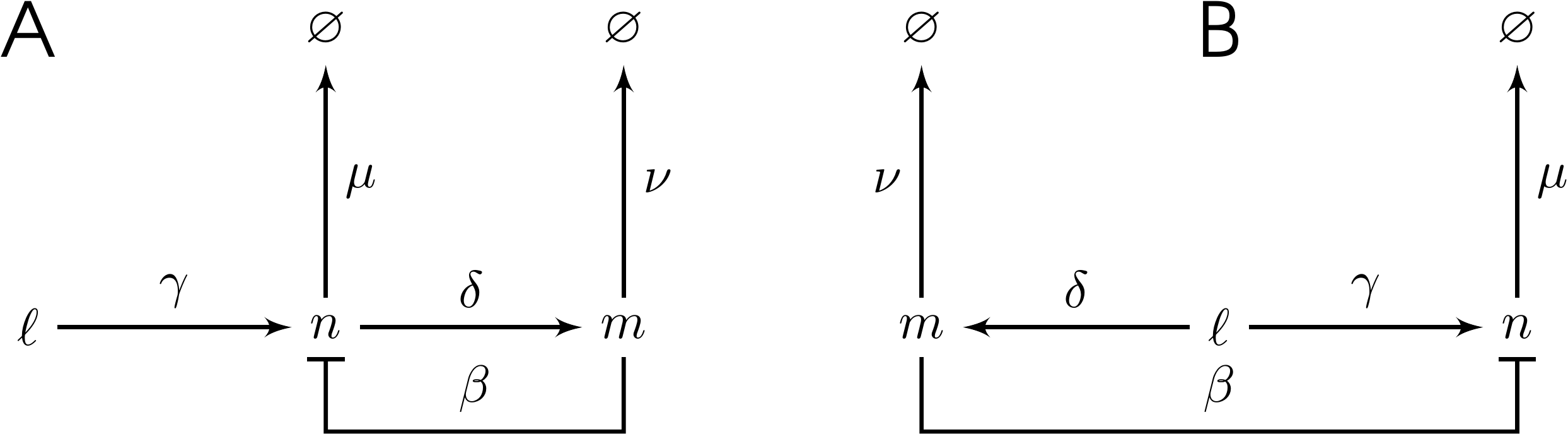}
		\caption{\label{fig:diagrams}
			feedforward (B) loops can generate non-monotonous kernels for the
			response of the output $n$ to the input $\ell$.} \end{center}
\end{figure}

The optimal kernel Eq.~\ref{eq:khonoise} for the noisy NM signal has a
complicated dependence on the parameters, but a simple time dependence
$k_*(t)\propto a^{+} e^{-\mu_*^{+} t}+a^{-}e^{-\mu_*^{-} t}$. In the interesting
underdamped regime, $\mu^{*\pm}=\mu_*\pm i\omega_*$ are complex conjugates
with positive real part, so that $k(t)\propto e^{-\mu_* t}[a \sin(\omega_* t) +
b \cos(\omega_* t)]$. Can such a damped oscillatory optimal kernel be
implemented with a simple two-stage biochemical network? As candidates we
consider incoherent feedforward and negative feedback networks
(Fig.~\ref{fig:diagrams}).

First, the (deterministic) rate equations
\begin{equation}
	\dot n = \gamma \ell - \beta m - \mu n ;\; \dot m = \delta n -\nu m
\end{equation}
describe a simple negative feedback on the output $n$, driven by the input
$\ell$. The feedback term is independent of $\ell$. This minimal, generic model
applies to a wide range of systems: circadian clocks based on negative feedback
and delay in gene expression~\cite{kondo97}, or negative feedback in signal
transduction pathways, like the well-characterized MAPK pathway
\cite{kholodenko00,sasagawa05}.

The response kernel can be found straightforwardly by expanding around the
steady state and Fourier transforming; one obtains
\begin{equation}
	k(t)\propto e^{-\frac{1}{2} t (\mu +\nu )} \bigl(\tfrac{\nu -\mu}{\beta'} \sin
	\bigl[\tfrac{t \beta'}{2}\bigr]+ \cos \bigl[\tfrac{t \beta'}{2}\bigr]\bigr),
\end{equation}
where $\beta'=\sqrt{4\beta \delta - (\mu -\nu)^2 }$ is real for sufficient
feedback and sufficiently matched time scales. Thus, negative feedback can
indeed generate damped oscillatory kernels.

Second, the rate equations
\begin{equation}
	\dot n = \gamma \ell - \mu n - \beta m;\; \dot m = \delta \ell -\nu m
\end{equation}
describe an incoherent feedforward loop, which is an omnipresent network motif in
cell signaling~\cite{alon06}. The response kernel here is
\begin{equation}
	k(t)\propto
	\frac{e^{-\mu t}
		[\beta \delta + \gamma(\mu-\nu)] -
		\beta\delta e^{-\nu t}}{\mu -\nu },
\end{equation}
which exhibits a single undershoot but no oscillatory regime. Such a kernel may
be optimal for predicting Markovian and non-Markovian signals that are well
above the noise floor, but cannot be optimal for noisy and strongly oscillatory
signals, since these require oscillatory kernels, as discussed above.

In summary, a simple negative feedback on the output production can produce
kernel shapes that are optimally predictive for stochastic NM signals; the
enveloping decay rate $(\mu+\nu)/2$ is the average decay rate, and the
oscillation period and phase are controlled by the decay rate mismatch and
feedback strength. These have to be matched to the signal to optimize the
predictive response. For example, in the noise-dominated and underdamped
regime, $\mu^{*\pm} \simeq (\eta \pm \eta')/2$ where $\eta$ is the signal
damping parameter. Matching $\beta' \simeq \eta'=\sqrt{4-\eta^2}$ then yields
the correct oscillation period for the kernel; the average decay rate should
match $(\mu+\nu)/2\simeq\eta$; and the phase shift should be adapted according
to the prediction interval. In contrast, a simple incoherent feedforward on the
production of the output does not allow for an oscillatory kernel, making this
motif less suited for predicting noisy oscillatory signals.

\subsection{A minimal signaling module}

To validate our linear theory in a simple concrete example, and to extend it to
finite molecule numbers, nonlinear response, and nonzero intrinsic noise, we
now consider a minimal sensory module in detail.

\begin{figure}[tb]
	\begin{center}
		\includegraphics[width=\columnwidth]{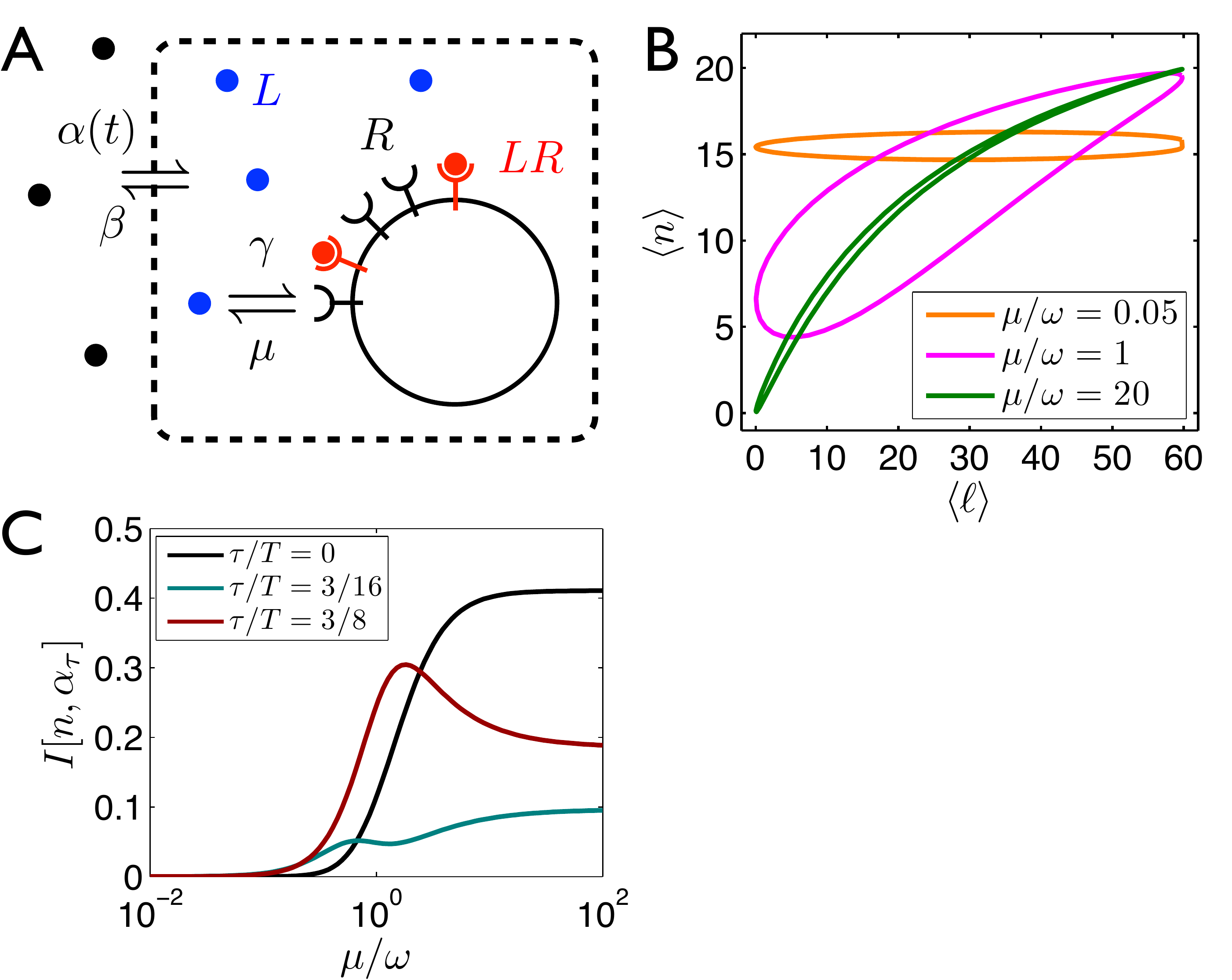}
		\caption{\label{fig:simpmodule}%
			Predictive information in the module Eq.~\ref{eq:reactions}. (A)
			Schematic of the module. The response network consists of the equilibrium
			binding reaction only, cf.~\ref{fig:linearintro}A. (B) Hysteresis in the
			response of the module to sinusoidal input $\alpha(t)=s(t)$. Note that in
			the saturating regime, the response is nonlinear. (C) Predictive
			information $\rho=0.5,\nu_0=10$ in the high-copy number limit,
			Eq.~\ref{eq:Ismalldriving}}
	\end{center}
\end{figure}

In this system (Fig.~\ref{fig:simpmodule}A), ligand molecules \lig~enter a
reaction volume at a periodically varying rate $s(t)=\alpha(t)$ proportional to
the background ligand concentration, which constitutes the input signal; they
are removed with rate $\beta$. A pool of \nmax~receptor molecules $\rec$ bind
ligands:
\begin{equation}
\label{eq:reactions}
\varnothing \underset{\beta}{\overset{\alpha(t)}{\rightleftharpoons}} \lig;\quad
\lig + \rec \underset{\mu}{\overset{\gamma}{\rightleftharpoons}}  \lr.
\end{equation}
The fast rates $\alpha(t)$ and $\beta$ dynamically create $\ell(t)$ free ligand
molecules, which act as noisy reporters for the input. The ligands drive the
formation of $x(t)=n(t)$ LR complexes, which form the output of the module,
with a lag depending on the unbinding rate $\mu$, Fig.~\ref{fig:simpmodule}B.
The module's stochastic dynamics are governed by the Master equation
\begin{equation} \label{eq:CME}
\partial_t p(\ell,n|t)
 = \bigl(\B_\ell^{\alpha(t),\beta} + \A_{\ell,n}^{\gamma,\mu,\nmax}\bigr)
 p(\ell,n|t),
\end{equation}
where $\B_\ell^{\alpha,\beta} = \alpha(\Em_\ell-1)+\beta(\Ep_\ell-1)\ell$ and
$\A_{\ell,n}^{\gamma,\mu,\nmax} = \gamma(\Ep_\ell\Em_n-1)\ell r
+\mu(\Em_\ell\Ep_n-1)n$ define the ligand exchange and ligand binding,
respectively, and $\Epm_x f(x) = f(x\pm1)$ are step operators. Appropriate
boundary conditions enforce $\ell\ge0$ and $0\le n\le \nmax$.

Eq.~\ref{eq:CME} yields the predictive information $I(\tau) =
I[n(t),\alpha(t+\tau)]$ without resorting to a Gaussian approximation. In a
linear regime and for high ligand numbers, as detailed in the subsection
\emph{Analytic results for predictive information} below, we obtain
\begin{equation}
\label{eq:Ismalldriving}
I(\tau) = \tfrac{1}{4} \rho^2\nu_0\delta^2 \cos^2[(\tau+\Lambda)\omega]
+O(\rho^4),
\end{equation}
where $\Lambda=\omega^{-1}\tan^{-1}(\omega/\mu)$ is the response lag time. As
shown in Fig.~\ref{fig:simpmodule}C the module exhibits lagging optimal
prediction above a critical $\tau_c=\cos^{-1}1/3\simeq0.2T$, exploiting
anticorrelated past signals $\tau+\Lambda\simeq T/2$ before the prediction
target; this is the phase-matching strategy found already for exponential
responders, Fig.~\ref{fig:optkernels}A, as explained next.

\subsubsection{Exploiting signal correlations through memory}

\begin{figure}[tb]
	\begin{center}
		\includegraphics[width=\columnwidth]{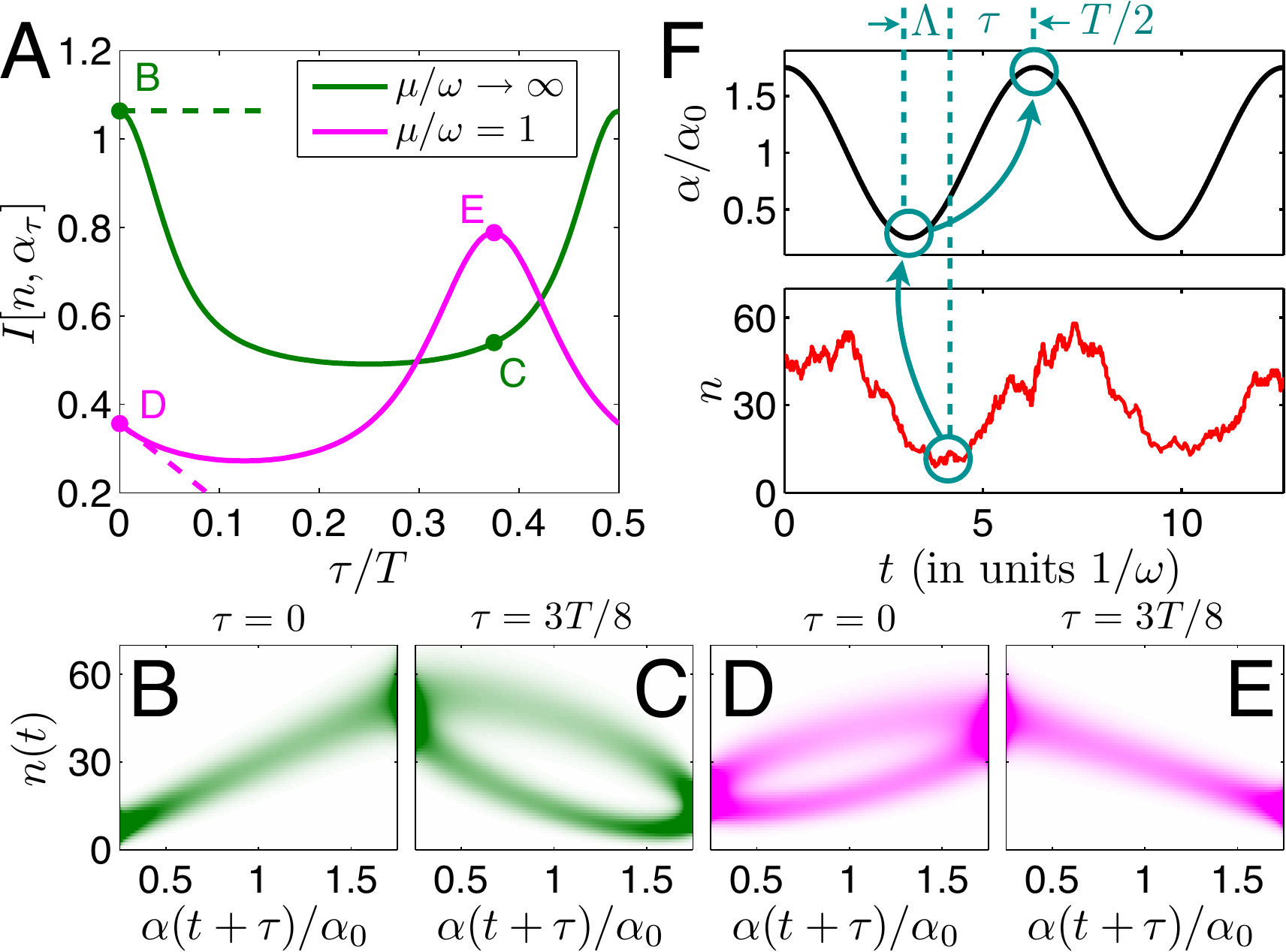}
		\caption{\label{fig:explanation}%
			Prediction through memory.
			(A) For a range of prediction intervals $\tau$, frequency matching
			$\mu=\omega$ is more predictive than perfect tracking $\mu\to\infty$.
			The reason, as demonstrated by $p(n,\alpha_\tau)$ at key points (B-E as
			indicated in A) and illustrated in F, is memory: receptors with $\mu=\omega$
			time-convolve the signal, so that $n(t)$ lags by $\Lambda$; it is maximally
			correlated with $\alpha(t-\Lambda)$, which in turn mirrors
			$\alpha(t-\Lambda+T/2)$.
			Parameters are $\rho = 0.75$ and $\nu_0 = 30$.
		}
	\end{center}
\end{figure}

The reason why a lagging response helps long-term prediction is that it removes
an ambiguity inherent the oscillatory signal, Fig.~\ref{fig:explanation}B-E. This
mechanism is a realization of the general mechanism that an optimal response
kernel emphasizes the most informative past signal values, in the particular
case of a deterministic sinusoidal input signal and an exponential response,
as in the minimal module Eq.~\ref{eq:reactions}.

While the fast response tracks the present input (B), its predictions suffer
from a two-fold ambiguity about $\alpha(t+\tau)$, since a given value of $n(t)$
maps with high probability to two distinct values of $\alpha(t+\tau)$ whenever
$\tau$ is not a multiple of $T/2$ (C). Indeed, the autocorrelation of a perfect
sinusoidal input is $r_\alpha(t)=\cos(t/T)$, showing extrema at multiples of
$T/2$. The frequency-matched response tracks the delayed signal
$\alpha(t-\Lambda)$, since an exponential responder effectively delays a
sinusoidal input signal by a constant shift $\Lambda$; this is a peculiarity of
this signal/responder combination. This delayed tracking introduces ambiguity
about the present signal for the same reason (D) but strikingly, helps
prediction: The future signal $\alpha(t-\Lambda+T/2)$ is tracked without
two-fold ambiguity (E), which maximizes predictive information. Indeed
Eq.~\ref{eq:Ismalldriving} shows that $I$ is maximal when the lag, advanced by
half a period, equals the prediction interval, $-\Lambda + T/2=\tau$. For
$\mu=\omega$, the lag is $\Lambda=T/8$ and the maximum is at $\tau=3T/8$.
Counter-intuitively, then a frequency-matched responder contains even more
information about the future ($\tau=3T/8$) than about the present,
Fig.~\ref{fig:explanation}A. The effect of disambiguation is strong enough to
outweigh the disadvantage of response damping ($I\propto \delta^2$, compare the
ranges in C and E).

\subsubsection{Details on fast ligand exchange}

To identify the conditions for which ligand fluxes dominate over binding fluxes,
we sum Eq.~\ref{eq:CME} over $n$, obtaining (here suppressing $t$ arguments)
\begin{multline}\label{eq:ellmarginal}
\dot p(\ell) = \B_\ell^{\alpha,\beta} p(\ell)
+ \gamma[\E{r|\ell_+}\ell_+p(\ell_+) - \E{r|\ell}\ell p(\ell)] \\
+ \mu[\E{n|\ell_-}p(\ell_-) - \E{n|\ell}p(\ell)].
\end{multline}
For moderate driving amplitude around half-filling of the receptors, the
conditional averages $\E{n|\ell}, \E{r|\ell}$ remain close to their mean
$\nu_0$. Therefore, taking $\beta\gg\gamma \nu_0$ and $\alpha_0\gg\mu\nu_0$
ensures that $\dot p(\ell)\simeq \B_\ell p(\ell)$, \textit{i.e.}~the ligand in-
and outflux terms dominate.

\subsubsection{High copy-number limit}

To gain intuition about the predictive performance of the module
Eq.~\ref{eq:reactions}, we consider a high copy-number limit. We let
$\{\lambda_0, \nmax\}\to\infty$ and $\gamma\to 0$ such that the mean number of
bound receptors $\nu_0 = \gamma\lambda_0 N/\mu$ remains constant. This ensures
that $n(t)$ remains in the linear, non-saturated range of the response curve:
$\nmax\gg\nu_0$ implies that $r(t)=O(\nmax - \nu_0) = O(\nmax)$. It also
ensures that the receptors are effectively driven by a deterministic signal:
The relative width of the ligand distribution decreases as
$\sigma_\ell/\lambda_0= O(\lambda_0)^{-1/2}$. The receptor dynamics thus reduce
to a birth-death process, $\partial_t p(n|t) = \B_n^{\g(t), \mu}p(n|t)$, with
effective birth rate $\gamma\ell(t) r(t) = \gamma [s(t)/\beta]N +
O(\nu_0/\nmax) + O(\lambda_0^{-1/2})$, which we denote as $\g(t)\equiv \gamma
[s(t)/\beta]N = \nu_0\mu[1+\rho\cos(\omega t)]$. The solution to a birth-death
process with time-dependent birth rate $\g(t)$ is a Poisson distribution
\cite{Mugler2010} with mean
\begin{equation}\label{eq:mean_n}
	\E{n(t)} = \int_{-\infty}^0 \,dt' e^{-\mu(t-t')} \g(t')
	= \nu_0\{1+\rho\delta\cos[\omega( t-\Lambda)]\},
\end{equation}
where the lag $\Lambda \equiv \tan^{-1}(\omega/\mu)/\omega$ and the damping
$\delta \equiv [1+(\omega/\mu)^2]^{-1/2}$, as in the main text.

\subsubsection{Analytic results for predictive information}

For a deterministic signal $\alpha(t)$, we have $p(\alpha_\tau|n,t)=p(\alpha_
\tau|t)=\delta[\alpha_\tau-\alpha(t+\tau)]$, and the predictive information
becomes
\begin{align}
	&I[n,\alpha_\tau] = \nonumber\\
	&= \sum_n \int
	d\alpha_\tau\ p(n,\alpha_\tau)\log\Bigl[\frac{p(n|\alpha_\tau)}{p(n)}\Bigr]
	\nonumber\\
	&= \sum_n \int d\alpha_\tau \oint
	dt\ p(\alpha_\tau|n,t)p(n|t)p(t)
	\log\Bigl[\frac{p(n|\alpha_\tau)}{p(n)}\Bigr]\nonumber\\
	&= \sum_n \frac{1}{T} \oint dt\ p(n|t)
	\log\Bigl[\frac{p(n|\alpha(t+\tau))}{p(n)}\Bigr].
\end{align}
For cosine
driving $\alpha(t)=\alpha(T-t)$, there is a two-to-one relationship between $t$
and $\alpha$.  This yields $p(n|\alpha(t+\tau))=[p(n|t_1)+p(n|t_2)]/ 2$, where
$t_1=t$ and $t_2=T-t-2\tau$ are the two time points for which $ \alpha_\tau$
takes on the value $\alpha(t+\tau)$.  The predictive information becomes
\begin{equation}
	\label{eq:Isimp} I[n,\alpha_\tau] = \sum_n \frac{1}{T} \oint dt\ p(n|t)
	\log\Bigl[\frac{p(n|t_1)+p(n|t_2)}{2p(n)}\Bigr].
\end{equation}
In the high copy-number limit (Figs.~\ref{fig:simpmodule}
and~\ref{fig:explanation}), Eq.~\ref{eq:Isimp} is evaluated numerically using
the Poissonian $p(n|t)$ with mean $\E{n(t)} = \nu_0\{1+\rho\delta\cos[\omega(
t-\Lambda)]\}$, time lag $\Lambda \equiv \tan^{-1}(\omega/\mu)/\omega$, and
damping factor $ \delta \equiv [1+(\omega/\mu)^2]^{-1/2}$.

To get analytical insight, we can expand Eq.~\ref{eq:Isimp} in the limit of small
driving amplitude $\rho$. To facilitate the expansion, we exploit the fact that
$p(n|t) $ can be expressed~\cite{Mugler2010} in terms of its Fourier modes in
$t$,
\begin{equation}
\label{eq:FT}
p(n|t) = \sum_{z=-\infty}^\infty p_n^z e^{-iz\omega t},
\end{equation}
and its natural eigenmodes in $n$,
\begin{equation}
\label{eq:pnz}
p_n^z = e^{iz\omega\Lambda}\sum_{j=0}^\infty
	\frac{(\nu_0\rho\delta/2)^{2j+|z|}}{j!(j+|z|)!}\phi_n^{2j+|z|}.
\end{equation}
Here $p_n^z=(1/T)\int_0^T dt\,e^{iz\omega t}p(n|t)$ are the components of the
Fourier transform, which have support only at integer multiples $z$ of the
driving frequency, and $\phi_n^j$ are the eigenmodes of the static birth-death
process with mean bound receptor number $\nu_0$, i.e. $-\B_n^{\nu_0,1}\phi_n^j=j
\phi_n^j$ for eigenvalues $j\in\{0,1,\dots,\infty\}$~\cite{Walczak2009}.
Eq.~\ref{eq:pnz} shows directly that the distribution is expressible as an expansion
in the small parameter $\rho$. The remaining task is then to identify the
leading term in $\rho$.

To identify the leading term in $\rho$, we insert Eq.~\ref{eq:FT} into
Eq.~\ref{eq:Isimp},
which yields
\begin{eqnarray}
\label{eq:I5}
I &=& \frac{1}{T}\sum_n\int_0^T dt\ \sum_z p_n^z e^{-iz\omega t}\nonumber\\
&&	\times\log \left\{ \frac{1}{2p_n^0}\sum_{z'}p_n^{z'}
	\left[e^{-iz'\omega t}+e^{-iz'\omega (T-2\tau-t)}\right]\right\}.\qquad
\end{eqnarray}
Here we have recognized that $p(n)$, which is the time average of $p(n|t)$, is
also the zeroth Fourier mode:
$p(n)=\int_0^Tdt\ p(n|t)p(t)=(1/T)\int_0^Tdt\ p(n|t)=p_n^0$.
Isolating the $z'=0$ term and defining $q_n^z\equiv p_n^ze^{iz\omega\tau}$ to
make the expression more symmetric yields
\begin{eqnarray}
I &=& \frac{1}{T}\sum_n\int_0^T dt\ \sum_z p_n^z e^{-iz\omega t}\nonumber\\
\label{eq:I6}
&&	\times\log \left\{1+\frac{1}{2p_n^0}\sum_{z'\neq0}q_n^{z'}
	\left[e^{-iz'\omega(t+\tau)}+e^{iz'\omega(t+\tau)}\right]\right\},\qquad
\end{eqnarray}
where we have recognized that $e^{-iz'\omega T}=1$.  Then, recognizing that the
term in brackets is symmetric upon $z'\to-z'$, we write the $z'$ sum in
terms of only positive integers,
\begin{eqnarray}
I &=& \frac{1}{T}\sum_n\int_0^T dt\ \sum_z p_n^z e^{-iz\omega t}\nonumber\\
\label{eq:I7}
&&	\times\log \left\{1+\frac{1}{2p_n^0}\sum_{z'>0}r_n^{z'}
	\left[e^{-iz'\omega(t+\tau)}+e^{iz'\omega(t+\tau)}\right]\right\},\qquad
\end{eqnarray}
where
\begin{eqnarray}
r_n^z &\equiv& q_n^z+q_n^{-z}\nonumber\\
\label{eq:rp}
&=& e^{iz\omega\tau}p_n^z+e^{-iz\omega\tau}p_n^{-z}\nonumber\\
&=& \left[e^{iz\omega(\Lambda+\tau)}+e^{-iz\omega(\Lambda+\tau)}\right]
	\sum_j \frac{(\nu_0\rho\delta/2)^{2j+|z|}}{j!(j+|z|)!}\phi_n^{2j+|z|}
\nonumber\\
\label{eq:rnz}
&=& 2\cos[z\omega(\Lambda+\tau)]
	\sum_j \frac{(\nu_0\rho\delta/2)^{2j+|z|}}{j!(j+|z|)!}\phi_n^{2j+|z|}
\end{eqnarray}
is a real quantity.

Now, since $r_n^z$ is expressed in terms of our small
parameter $\rho$, we Taylor expand the log in Eq.~\ref{eq:I7}:
\begin{eqnarray}
I &=& \frac{1}{T}\sum_n\int_0^T dt\ \sum_z p_n^z e^{-iz\omega t}
	\sum_{k=1}^\infty\frac{(-1)^{k+1}}{k}\nonumber\\
\label{eq:I8}
&&	\times\left\{\frac{1}{2p_n^0}\sum_{z'>0}r_n^{z'}
	\left[e^{-iz'\omega(t+\tau)}+e^{iz'\omega(t+\tau)}\right]\right\}^k.\quad
\end{eqnarray}
It will turn out that the first two terms in the Taylor expansion will
contribute to the leading order in $\rho$.  The first term ($k=1$) is
\begin{eqnarray}
I^{(1)} &=& \sum_n	\frac{1}{2p_n^0}\sum_{z'>0}r_n^{z'}
	\sum_z p_n^z\nonumber\\
\label{eq:I9}
&&	\times\frac{1}{T}\int_0^T dt\ e^{-iz\omega t}
	\left[e^{-iz'\omega(t+\tau)}+e^{iz'\omega(t+\tau)}\right],\qquad
\end{eqnarray}
where we have reordered terms in preparation for exploiting the relation
$(1/T)\int_0^T dt\ e^{-i(z-z')\omega t}=\delta_{zz'}$.  This relation turns the
two terms in brackets into Kronecker deltas, which each collapse the sum over
$z$, leaving
\begin{eqnarray}
I^{(1)} &=& \sum_n \frac{1}{2p_n^0}\sum_{z'>0}r_n^{z'}
	\left(e^{-iz'\omega\tau}p_n^{-z'}+e^{iz'\omega\tau}p_n^{z'}\right)
	\nonumber\\
\label{eq:I11}
&=& \frac{1}{2}\sum_n\frac{1}{p_n^0}\sum_{z'>0}(r_n^{z'})^2.
\end{eqnarray}
In a completely analogous way, the second term in the Taylor expansion ($k=2$)
reduces to
\begin{equation}
\label{eq:I12}
I^{(2)} = -\frac{1}{8}\sum_n \frac{1}{\left(p_n^0\right)^2}
	\sum_{x>0}\sum_{y>0}r_n^{x}r_n^{y}
	\left(r_n^{x+y}+r_n^{x-y}\right).
\end{equation}
Considering the $j=0$ term in $r_n^z$ (Eq.~\ref{eq:rnz}), it is clear that the leading
order behavior in $\rho$, proportional to $\rho^2$, comes from the $z'=1$ term
in Eq.~\ref{eq:I11} and the $x=y=1$ term in Eq.~\ref{eq:I12}:
\begin{eqnarray}
\label{eq:I13}
I &\approx& \frac{1}{2}\sum_n \frac{1}{p_n^0}(r_n^1)^2
	-\frac{1}{8}\sum_n \frac{1}{\left(p_n^0\right)^2}
	r_n^{1}r_n^{1}\left(r_n^{0}\right)\\
\label{eq:I14}
&=& \frac{1}{2}\sum_n \frac{(r_n^1)^2}{r_n^0}\\
\label{eq:I15}
&\approx& \cos^2[\omega(\Lambda+\tau)]
	\left(\frac{\nu_0\rho\delta}{2}\right)^2
	\sum_n\frac{(\phi_n^1)^2}{\phi_n^0}\\
\label{eq:I16}
&=& \frac{\nu_0^2\rho^2}{4}
	\left[ \frac{\cos(\omega\tau) - (\omega/\mu)\sin(\omega\tau)}
	{1+(\omega/\mu)^2} \right]^2
	\sum_n\frac{(\phi_n^1)^2}{\phi_n^0}.\qquad
\end{eqnarray}
Here, Eq.~\ref{eq:I14} uses the fact that $r_n^0=2p_n^0$ (Eq.~\ref{eq:rp}),
Eq.~\ref{eq:I15} takes
only the $j=0$ term in Eq.~\ref{eq:rnz}, and Eq.~\ref{eq:I16} recalls that
$\delta = [1+ (\omega/\mu)^2]^{-1/2}$ and uses
\begin{eqnarray}
	\cos[\omega(\Lambda+\tau)]
	&=& \cos(\omega\Lambda)\cos(\omega\tau)-\sin(\omega\Lambda)\sin(\omega\tau)
	\nonumber\\
	&=& \cos\left[\tan^{-1}(\omega/\mu)\right]\cos(\omega\tau)\nonumber\\
	&&	-\sin\left[\tan^{-1}(\omega/\mu)\right]\sin(\omega\tau)\nonumber\\
	&=& \frac{1}{\sqrt{1+(\omega/\mu)^2}}\cos(\omega\tau)\nonumber\\
	\label{eq:cosprop}
	&&	-\frac{\omega/\mu}{\sqrt{1+(\omega/\mu)^2}}\sin(\omega\tau).
\end{eqnarray}
The sum in Eq.~\ref{eq:I16} is evaluated by noting that the zeroth eigenmode is a
Poisson distribution with mean $\nu_0$ and that the first eigenmode is related
to the zeroth eigenmode via $\phi_n^1 = \phi_{n-1}^0-\phi_n^0 = \phi_n^0(n-
\nu_0)/\nu_0$~\cite{Mugler2009}.  The sum therefore becomes $(1/\nu_0^2)\sum_n
\phi_n^0(n-\nu_0)^2$, which is the variance of the Poisson distribution (equal
to $\nu_0$) divided by $\nu_0^2$, or $1/\nu_0$.  Altogether, then, Eq.~\ref{eq:I16}
becomes
\begin{equation}
\label{eq:I17}
I = \frac{\nu_0\rho^2}{4}
{\left[ \frac{\cos(\omega\tau) - (\omega/\mu)\sin(\omega\tau)}
		{1+{(\omega/\mu)}^2} \right]}^2,
\end{equation}
as in Eq.~\ref{eq:Ismalldriving}.

\subsubsection{Robustness to low copy-number effects}

\begin{figure}[tb]
	\begin{center}
		\includegraphics[width=\columnwidth]{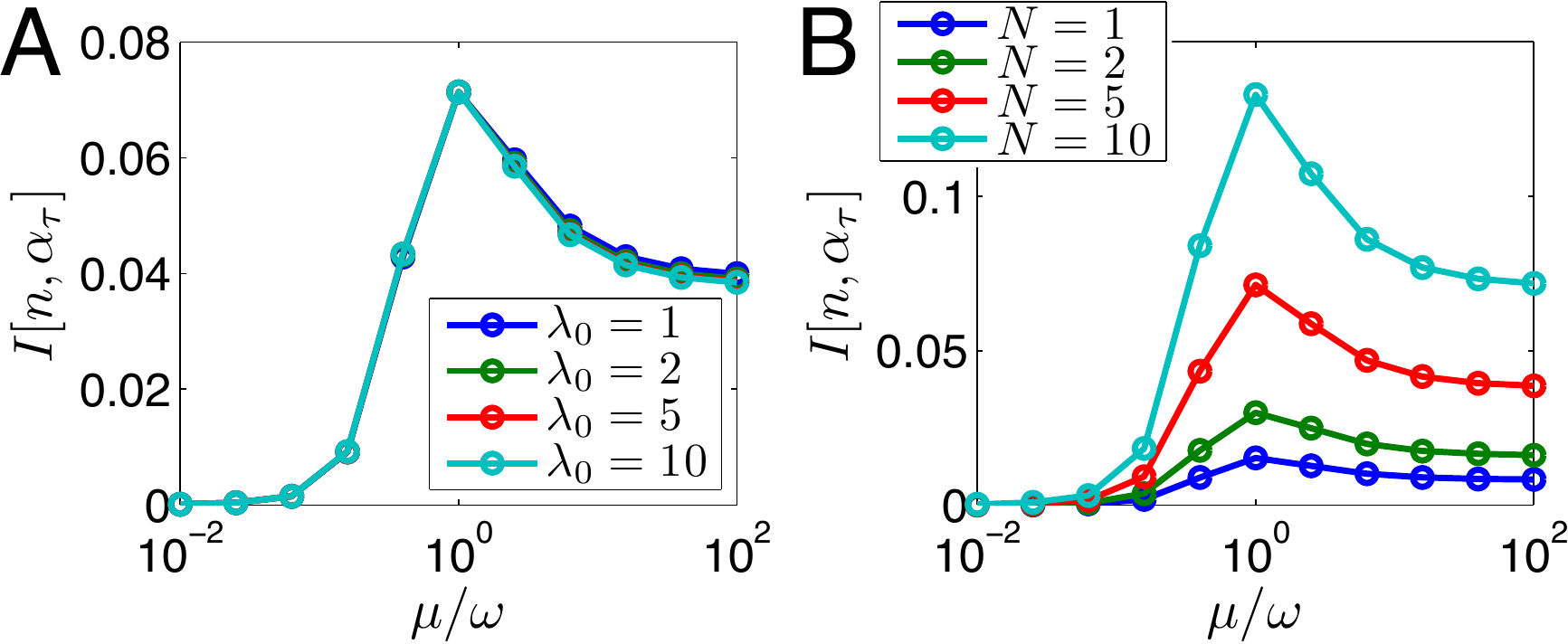}
		\caption{\label{fig:copynumber}%
			Lagging optimal prediction persists (A) for low mean ligand
			number $\lambda_0$ and (B) for low total receptor number $N$. The minimal
			module Eq.~\ref{eq:reactions} is driven by sinusoidal input $\alpha(t)$.
			Parameters are $\tau = 3T/8$, $\rho = 0.5$, $\beta/\omega = 100$, $\gamma =
			\mu/\lambda_0$, and, in A, $N=5$ and, in B, $\lambda_0 = 5$.  In A, all
			curves closely overlap.}
	\end{center}
\end{figure}

We here relax the assumption of high copy number and solve numerically the full
description of the system given by Eq.~\ref{eq:CME}.  We find that lagging
prediction remains optimal as the mean ligand number $\lambda_0$ and the total
receptor number $N$ are reduced, even down to $\lambda_0 = 1$
(Fig.~\ref{fig:copynumber}A) and $N = 1$ (Fig.~\ref{fig:copynumber}B).  As $N$
is reduced, the information is reduced for all values of the response rate
$\mu$ (B), since reducing $N$ compresses the response range.  As $\lambda_0$ is
reduced, the information is largely unchanged (A); this is because ligand
exchange remains faster than the driving dynamics ($\beta/\omega\gg 1$),
meaning that even a small number of ligand molecules can cycle in and out of
the system many times over a period.  In both cases, there remains an optimum
in the predictive information as a function of $\mu$ located in the regime $\mu
\simeq \omega$, illustrating that lagging optimal prediction persists even at
low copy numbers.

\subsubsection{Lagging optimal prediction of diverse input signals}

\begin{figure}[tb]
	\begin{center}
		\includegraphics[width=\columnwidth]{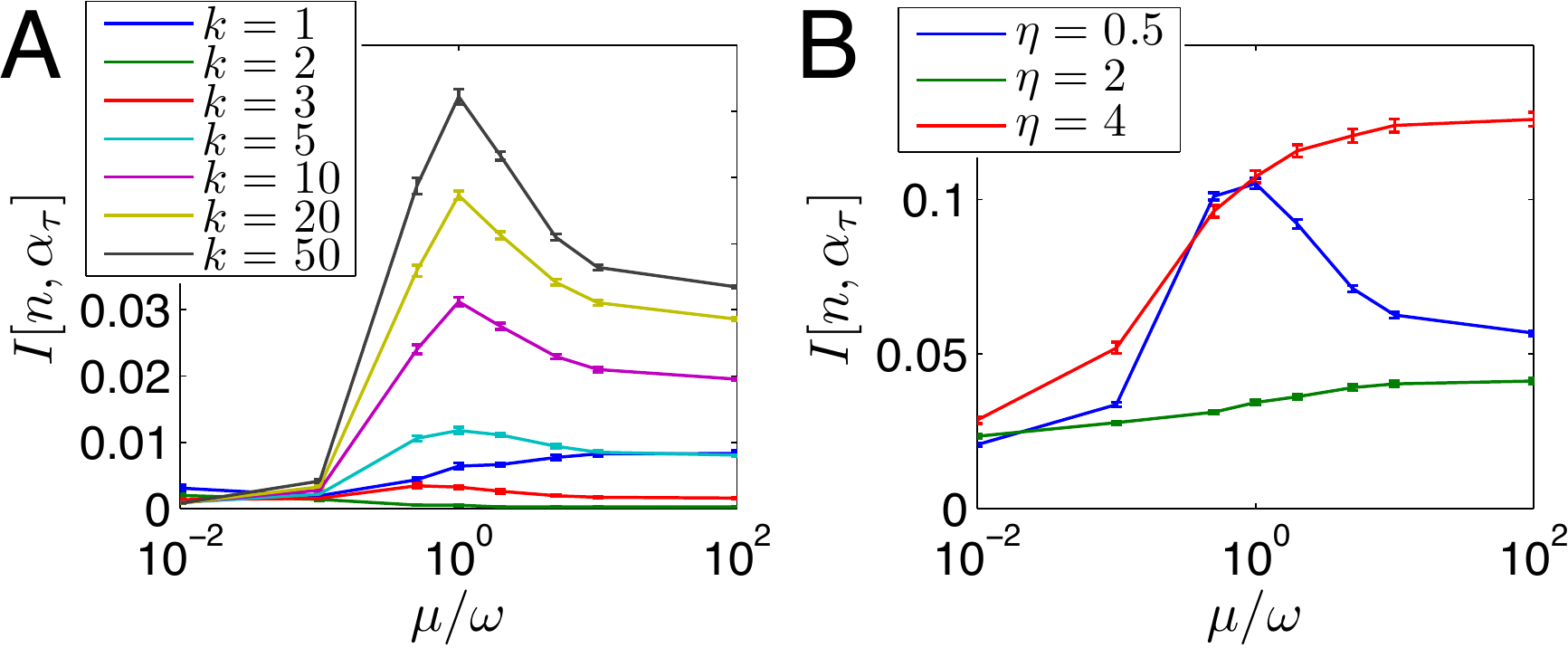}
		\caption{\label{fig:generic}%
			Lagging optimal prediction is generic.
			(A) For stochastic two-state driving with Gamma-distributed waiting times,
			lagging prediction is optimal for shape parameters $k\ge 3$. The mean
			waiting time is held constant at $T/2$.
			(B) For a noisy harmonic oscillator signal, fast prediction is optimal in the
			overdamped regime $\eta\geq2$. In the underdamped regime, lagging,
			frequency-matched prediction again becomes optimal. Parameters are $\tau =
			3T/8$, $\rho=0.5$, $\lambda_0=25$, $N=25$, $\beta/\omega=100$, and
			$\gamma=\mu/\lambda_0$. Simulation details are given in the last section of
			this supplement.}
	\end{center}
\end{figure}

Lagging prediction is a robust strategy for non-Markovian signals, beyond
Gaussian signals and linear response. Fig.~\ref{fig:generic}A shows the benefit
of a slow response for prediction for two-state switching signals $\alpha(t) =
\alpha_0 (1\pm \rho)$ with random switching times that are Gamma-distributed
with shape parameter $k$ and mean $T/2$. The minimal module's response strongly
distorts the rectangular signal shape, but lagging prediction remains optimal
for all $k\geq3$, Fig.~\ref{fig:generic}A.

Eq.~\ref{eq:Ismalldriving} demonstrates optimal lagging prediction for
non-Gaussian (deterministic) input at constant amplitude. The effect persists
also at low copy numbers. Fig.~\ref{fig:generic}B shows lagging optimal
prediction for $I[n(t),\alpha(t+\tau)]$ in the minimal module
Eq.~\ref{eq:reactions}, driven by NM signals as defined above, at $N=25$, but
the effect persists even for ligand or receptor numbers down to 5 (not shown).

\subsubsection{Ligand binding simulation parameters and details}

The data for Figs.~\ref{fig:generic}A and B were generated by a Gillespie-type
kinetic Monte Carlo simulation. Dynamic ligand birth rates $\alpha(t)$ were
approximated as constant during short discretization intervals of length
$\tau_\alpha=T/50$; after each such interval, queued next reaction times were
erased and re-generated according to the new value of the rate. This is an exact
simulation procedure for the approximated system with stepwise-constant rate.

For the two-state driving protocol (Fig.~\ref{fig:generic}A), the mean ligand
number and total receptor number were set to $\lambda_0=N=25$. The ligand death
rate was set to $\beta=100$, and the mean driving period was set to $T=2\pi$ in
simulation time units. Switching times were generated independently, following a
Gamma- (or Erlang-) distribution with shape parameter $k\in\{1,2,3,5,10,20,50\}$
and mean $T/2$. The input rate was set to a random initial value $\alpha_0[1\pm
\rho]$ and then toggled after each random switching time between $\alpha_0[1\pm
\rho]$, where $\alpha_0=\lambda_0\beta$ and $\rho=0.5$. For a given ligand
dissociation rate constant $\mu$, the association constant was set to
$\gamma=\mu/\lambda_0$ to ensure half-filling at the average driving rate.

For the harmonic-oscillator protocol (Fig.~\ref{fig:generic}B) the same
parameters were used, except that the driving signal was now generated by a
forward-Euler integration of the Langevin equation given in the main text. The
damping parameter $\eta$ was varied in $\{1/2,1,2,4\}$.

For each value of $\mu$, the system state was initialized to the equilibrium
molecule numbers at $\alpha=\alpha_0$, and $N_\mathrm{tr}=2000$ trajectories of
length $10T$ were generated.

Trajectories were sampled at discrete time intervals $T/100$, and the
corresponding samples of the input rate were binned. The input-output mutual
information was estimated by applying the definition $I[x,y]=\E{\log
\frac{p(x,y)}{p(x)p(y)}}$ to the binned simulation data. In doing so, the
choice of bin size for
the continuous variable $\alpha$ (in the harmonic-oscillator case) can lead to
systematic errors; we found the results for $I$ to be independent of the bin
size in a plateau region around $N_\mathrm{bin}=100$ equally filled bins, and
therefore used this binning for Fig.~\ref{fig:generic}B.

The data were split into 10 blocks of 200 trajectories each and the mutual
information $I[n,\alpha_\tau]$ for various prediction intervals $\tau\in [0, T/
2]$
were calculated based on histograms of the discrete-valued samples, for each
block. Plots show the averages over blocks together with standard errors of
the mean estimated from block-wise variation.

{
\subsection{\emph{E.\,coli} chemotaxis simulations}
\label{sub:ecoli}

To explore the role of prediction in chemotaxis, we set up a stochastic
simulation of \textit{E.\,coli} swimming in a concentration field in two
dimensions. \textit{E.\,coli} performs the well-known chemotaxis
behavior~\cite{berg04} of alternating runs and tumbles. During a tumble phase,
the bacterium remains stationary, reorients in a random direction and initiates
a new run with constant propensity $\beta=10/{\rm s}$. Runs then proceed with
constant velocity $v_0=20\mu {\rm m/s}$. They are governed by the Langevin
equation
\begin{align}
	\label{eq:ecolirun}
	d\vec r(t) &= v_0\, (\cos \phi, \sin \phi)\,dt, \nonumber\\
	d\phi(t) &= \sqrt{2D_\text{rot}}\,dw(t),
\end{align}
where $dw(t)$ is standard Gaussian white noise and affects the orientation
$\phi$ with a rotational diffusion constant
$D_\text{rot}=0.15\,\mathrm{rad^2/s}$.

\begin{figure}[tb]
	\centering
	\includegraphics[width=1.\columnwidth]{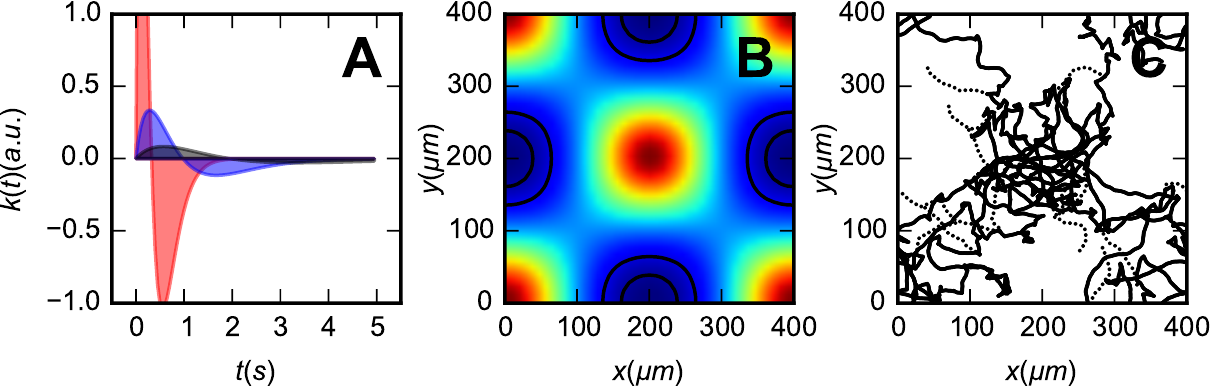}
	\caption{\label{fig:ecoli_intro}%
		Chemotaxis in \emph{E.\,coli}. (A) Linear chemotactic response kernel
		$k$ as measured in~\cite{segall86} (blue), and rescaled $k_5$ (red) and
		$k_{1/2}$ (black) versions. (B) Egg-crate potential, varying from 0 (blue)
		to 2 (red) in arbitrary units. (C) Example trajectory of a chemotaxing
		bacterium (solid line) with some virtual continued runs (dashed, see text).}
\end{figure}

Runs end with tumbling events, which have a propensity (tumbling rate)
\begin{equation}
	\label{eq:tumbleprop}
	\alpha(t) = \alpha_0\max[0, 1-x(t)];\;
	x(t)=\int_{-\infty}^{t} k(t-t') \ell(t')dt'.
\end{equation}
The base tumbling rate $\alpha_0=1/{\rm s}$. The tumbling rate is modulated as
a function of the signal history via the linear-response kernel $k(t)$, which
generates the chemotaxis pathway output $x(t)$. That is, the kernel $k$
summarizes the dynamics of the chemotaxis network including phosphorylation of
CheYp by CheA, modulation of CheA activity by the receptor, and
receptor-activity modulation via receptor-ligand binding and receptor
methylation and demethylation~\cite{shimizu10}, in a linear regime.

Following~\cite{celani10}, we use for $k(t)$ the clockwise-bias kernel as
measured on tethered bacteria~\cite{block82,segall86}. (Although this is not
fully rigorous, determining the actual rate kernel~\cite{block82} would require
additional assumptions, and tends to result in similar kernel shapes [not
shown]). For ease of use in the simulations, we fit the kernel to
$k(t)=be^{-at}(t-(a/2)t^2)$ where the data in~\cite{segall86} yield $b=A\times
2.91$ and $a=2.05/s$. The sensitivity $A=200/{\rm s}$ controls the degree of
tumbling rate modulation; we obtain a range of roughly $0<\alpha(t)\lesssim
3\alpha_0$ similar to what is reported in~\cite{segall86}.

Importantly, the kernel $k$ has the property of perfect adaptation
$\int_{0}^{\infty} k(t)dt=0$, which implies, among other advantages, that
it is insensitive to the constant background concentration level. As the kernel
shape suggests (Fig.~\ref{fig:ecoli_intro}A), the output is a smoothed
finite-time derivative of the input. Here, we do not attempt to derive a
globally optimal shape, but start with the observation that the kernel is
adaptive, and ask how the optimal speed of the kernel depends on the noise. We
do this by comparing rescaled kernels $k_\nu(t) = \nu^{2}k(\nu t)$ of varying
response speed $\nu$. With the chosen amplitude scaling, $\int
k_\nu(t-t')\ell(t')dt'\to \int |k(t)|dt \times \dot{\ell}(t)$ as
$\nu\to\infty$, \textit{i.e.}~in this limit the kernel acts as the instantaneous
input derivative. 

\begin{figure}[tb]
	\includegraphics[width=1.\columnwidth]{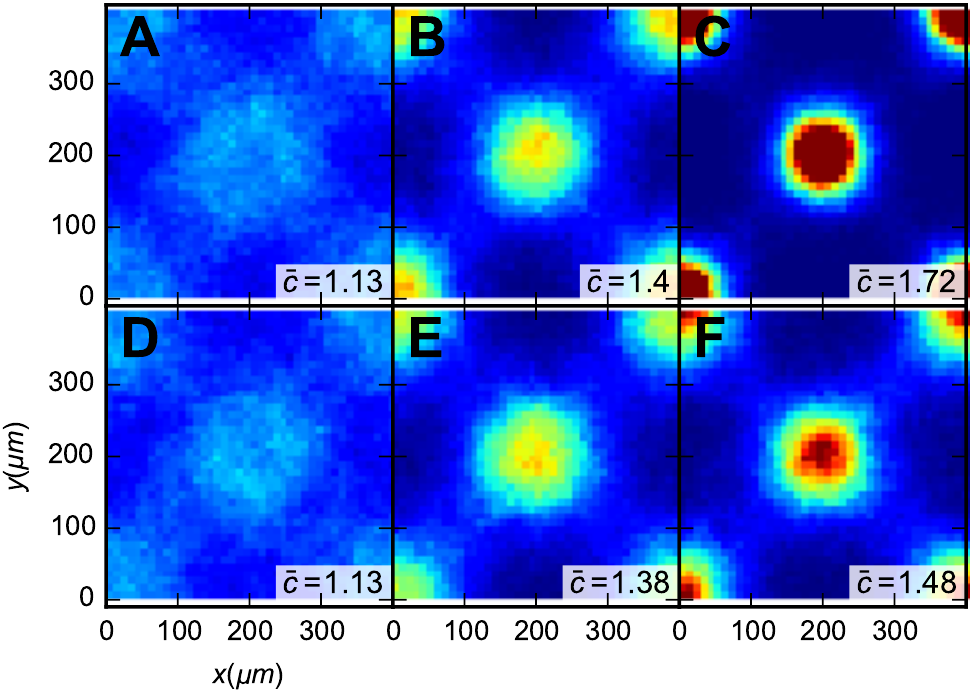}
	\caption{\label{fig:ecoli_hists}%
		Chemotaxis in an egg-crate potential. Average concentration $\bar c$ at
		noise $\theta=0$ (A-C) and at noise $\theta=2$ (D-F). The slow kernel
		$k_{1/2}$ (A,D) has a low performance due to a lack of response speed, but
		is robust to noise. The fast kernel $k_3$ (C,F) shows the highest
		localization, even though it is strongly affected by noise. The wild-type
		kernel $k_1$ (B,E) shows intermediate behavior.}
\end{figure}


Simulated \emph{E.\,coli} bacteria chemotax in a $L\times L$ concentration
field with periodic boundary conditions of size $L=400\mu\mathrm m$, with
concentrations given by a sinusoidal egg-crate function $c(x,y) = 1 + \cos(2\pi
x/L)\cos(2\pi y/L)$ ($c$ is unit-less; see Fig.~\ref{fig:ecoli_intro}B). The
signal $s(t)$ is the ligand concentration: $s(t) = c[\vec r(t)]$. In this
parameter regime, the resulting input signal is not oscillatory: the
correlation function decays monotonically (but not exponentially, not
shown).

To appreciate the role of input noise, we compare the case without noise in the
input signal, with the scenario where due to a low concentration of ligand, the
noise is dominated by the input noise, and the intrinsic noise can be
neglected, as in our theory. For the latter case, we now determine a reasonable
noise strength.
The input signal of the chemotaxis system is the activity $A\propto \ell$ of
the receptor molecules that bind the ligand. Input noise arises from
receptor-ligand binding and/or the conformational dynamics of the receptor
molecules. Since we disregard intrinsic noise, only the relative noise strength
matters, as well as its correlation time, as discussed previously.
If there are a total of $R_T$ independent receptor units that switch between an
active and an inactive state, then the relative variance $\sigma^2_A / \E{A}^2
= R_T p(1-p)/(R_T p)^2\simeq (R_T p)^{-1}$, where $p$ is the (small) average
activity of a receptor unit.
With a dissociation constant of $K_D \simeq 0.1-1 \mu{\rm M}$ \cite{vaknin07},
a concentration $c\simeq 1{\rm nM}$ yields a receptor occupancy (and we assume,
activity) of $p\approx 0.001-0.01$. We assume that $R_T$ is given by the number
of receptor-CheA complexes, yielding $R_T \approx 1000$~\cite{li04}. Moreover,
we assume that the correlation time is given by that of receptor-ligand
binding, $\tau_A = 1 / (k_{\rm on}c + k_{\rm off})$, where $k_{\rm on}$ and
$k_{\rm off}$ are the receptor-ligand association and dissociation rates,
respectively. These assumptions most likely give a lower bound on the noise,
because cooperative interactions between receptors~\cite{skoge13} and diffusion
of ligand~\cite{wang07} introduce spatio-temporal correlations between the
receptors. With a dissociation constant $K_D=0.1\mu{\rm M}$ and an association
rate of $k_{\rm on}=10^9 {\rm M}^{-1}{\rm s}^{-1}$ \cite{danielson94}, the
dissociation rate is $k_{\rm off}\approx 100 {\rm s}^{-1}$, which yields a
correlation time $\tau_A\approx 10 {\rm ms}$. We then arrive at an integrated
noise strength relative to the signal of $\sigma^2_A\tau_A / \E{A}^2 \approx
1-10{\rm ms}$.
The effect of noise on the network output is determined by the relative
integrated noise strength $\tau_A\sigma_A^2/\E{A}^2$. In our simulations, we
use a time step of $dt = 2 {\rm ms}$; to replicate the noise strength we thus
generate the network input $\ell=s+\xi$ by adding to the signal at each
time step an independent Gaussian random number $\xi$ of relative variance
$\theta^2\equiv\sigma_\xi^2/\E{s}^2 = (\tau_A/dt) / (pR_T) \approx 0.5 - 5$. As
a representative relative noise we have finally chosen $\theta=2$.

The simulation proceeds by random reorientation followed by Euler forward
integration of Eq.~\ref{eq:ecolirun} during run phases, and simultaneously,
updates of the pathway output $x(t)$ during run and tumble phases.
Tumble/restart events are generated via the procedure described in
Ref.~\cite{albada09a}. Fig.\ref{fig:ecoli_intro}C shows an example trajectory
segment.

As a result of the chemotactic mechanism, the simulated bacteria move on
average towards regions of higher input concentration, as indicated by the
long-term average concentration $\bar{c}=\E{s}$, which exceeds the random
value, $\bar{c}=1$, see Fig.~\ref{fig:ecoli_hists}.

\subsubsection{Coarse-grained description as biased diffusion}%
\label{ssub:coarse_grained_description_as_biased_diffusion}

The stationary distribution does not show the speed at which the concentration
maxima are approached by the bacteria. To measure the speed, we calculated the
average, coarse-grained chemotactic speed on the time scale $\delta t$ as
\begin{equation}
	\label{eq:csspeed}
	\vec v_{\delta t}(\vec r) =
	\E{\vec r(t+\delta t) - \vec r(t)}_{\vec r}/\delta t,
\end{equation}
where the subscript indicates that the average is taken over trajectories that
pass through $\vec r$ at $t$. Generally speaking, the choice
of mesoscopic time scale $\delta t$ affects the measured chemotactic speeds
when $\delta t$ is smaller than than a typical run length; however too large
$\delta t$ will include effects from the curvature of the underlying
concentration field. We chose $\delta t=4\mathrm s$ as a reasonable compromise;
results from 2s to 8s are similar.

\begin{figure}[tb]
	\includegraphics[width=1.\linewidth]{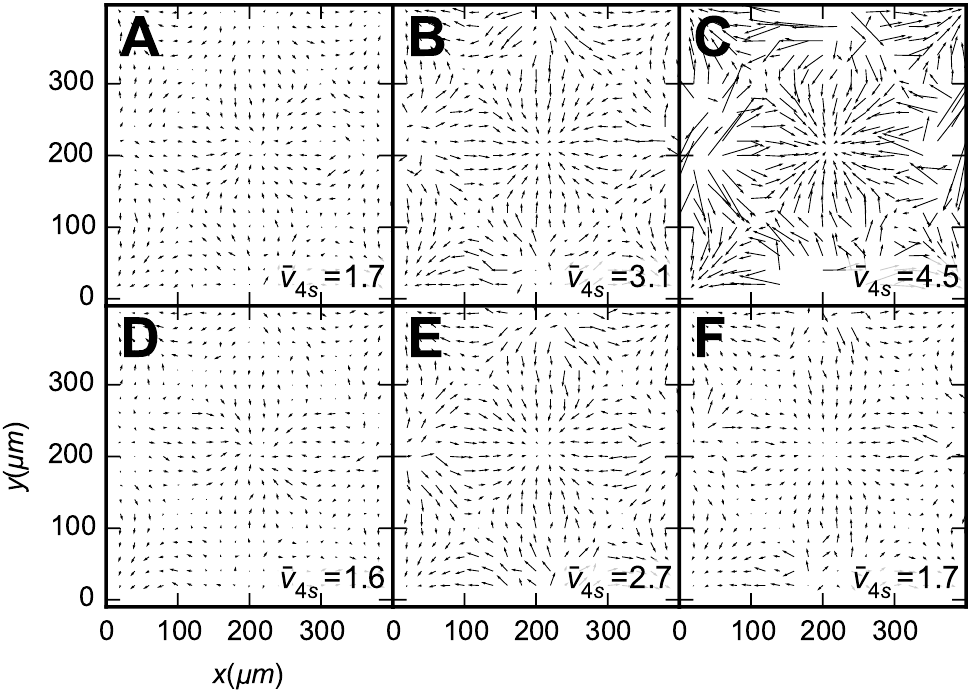}
	\centering
	\caption{\label{fig:speedfield}%
	Coarse-grained local velocity $v_{\delta t}$ in $\mu\mathrm m/\mathrm s$ with
	time-step $\delta t=4\mathrm s$. As in Fig.~\ref{fig:ecoli_hists}, noise levels are
	$\theta=0$ (A-C), $\theta=2$ (D-F), and kernel speeds are $\nu=1/2$
	(A,D), $\nu=1$ (B,E) and $\nu=3$ (C,F), respectively.}
\end{figure}

This vector field is shown in Fig.~\ref{fig:speedfield}. Clearly, at
$\theta=0$, the mean speed increases as the kernel integration time
decreases; however at finite noise $\theta=2$, there is a speed optimum for
the wildtype-speed kernel in panel (E). The average uphill speed, defined as
the average projection of $v_{\delta t}$ in the direction of the positive
concentration gradient,
\begin{equation}
	\label{eq:avuphill}
	\bar v_{\delta t} = \Bigl\langle\frac{\vec v_{\delta t}
		(\vec r)\cdot \nabla c(\vec r)}{\|\nabla c (\vec r)\|} \Bigr\rangle,
\end{equation}
shown in the insets, confirms this observation.

The question arises why localization efficiency at high noise stays high for
the fast kernel $k_3$ (Fig.~\ref{fig:ecoli_hists}), while the drift speed
towards the maxima actually decreases. This can be understood by considering
that both the chemotactic speed and the localization are affected by the
typical run length. Namely, if runs become shorter, the chemotactic speed
decreases because random reorientations occur more often. At the same time,
with shorter runs, smaller features of the concentration field can be resolved,
increasing the potential for good localization. Indeed we find that unfiltered
noise $\theta=2$ induces many tumbles for the fast kernel $k_3$, decreasing the
mean run length to 0.55s, whereas in all other conditions, the mean run length
remains close to the unperturbed value 1s. Thus, short runs allow slow but in a
static concentration field eventually successful, chemotaxis.

\subsubsection{Predictive information in chemotaxis}%
\label{ssub:predictive_information_in_chemotaxis}

To chemotax, bacteria modulate the run lengths. Since a tumbling rate bias at
the current time comes into effect only at the next tumbling event, this means
that the current modulation of the tumbling rate should take into account
the future ligand concentration until the end of the run, around $\alpha_0^{-1}
= 1{\rm s}$ in the future. In other words, \textit{E.\,coli} needs to predict
future ligand concentration. To elucidate what precise kind of predictive
information is most relevant for chemotaxis, and to relate it to chemotactic
performance, we measured predictive information in our simulations. 

Since \textit{E.\,coli} reacts to its own predictions by tumbling, correlating
the current signaling output with the future concentrations along its
trajectory would include the feedback loop between prediction and motile
response. 
To assess how reliably \textit{E.\,coli} can predict the future concentration,
we thus need to decouple the estimation of the future concentration
from the response to the estimate, which is the tumble event. We do this by 
constructing virtual trajectories (shown as dotted lines in
Fig.~\ref{fig:ecoli_intro}C), which are obtained by prolonging a run for a
set amount of time after a tumble event. Along these virtual runs
we then compute the predictive information between the output $x(t)$ at a
given point in time $t$, and either the future signal $s(t+\tau)$ or the future
change in signal $s(t+\tau) - s(t)$. It is this predictive information
that allows \textit{E.\,coli} to estimate the benefit of postponing or
advancing the next tumble event.

\begin{figure*}[t]
	\centering
	\includegraphics[width=1.\textwidth]{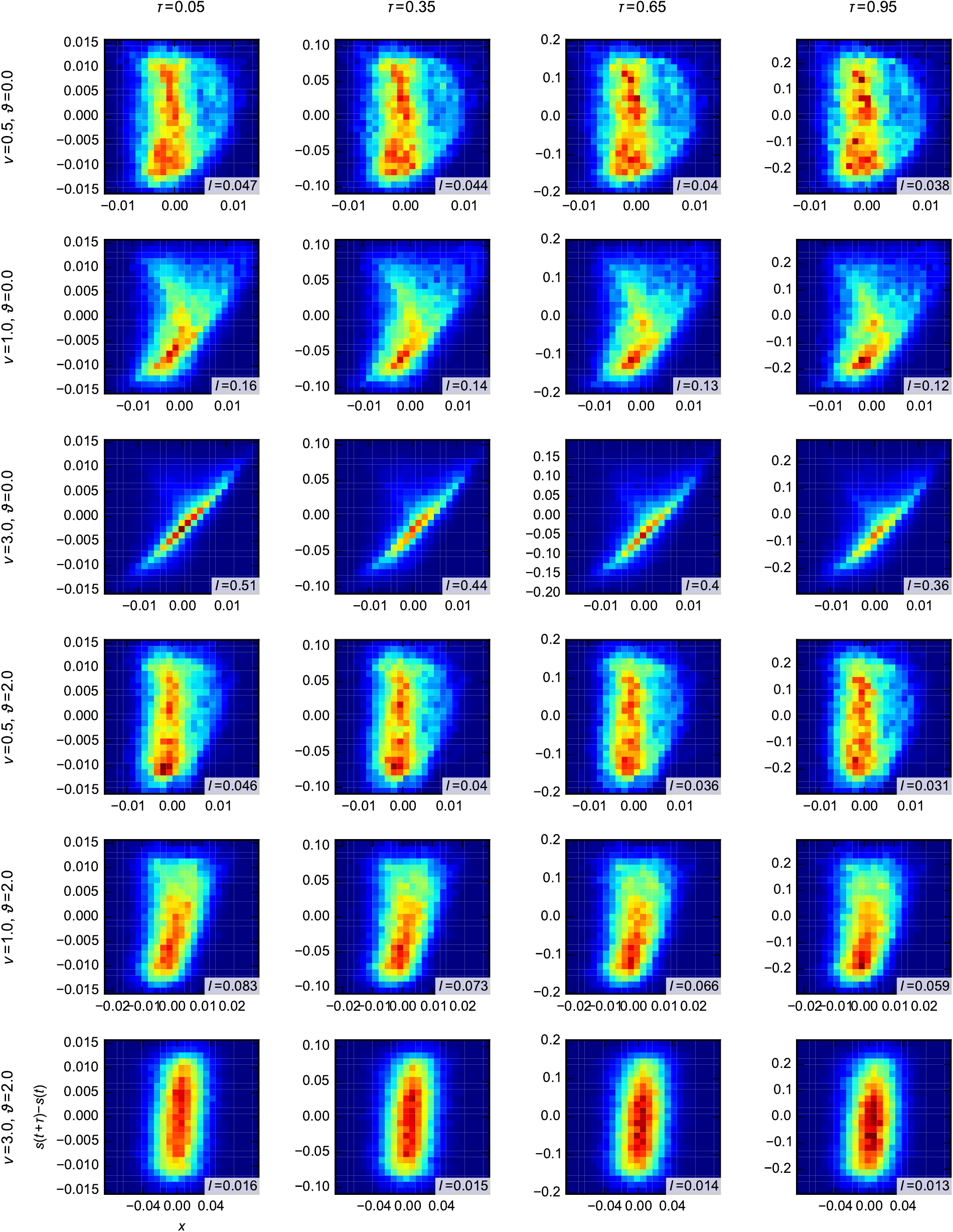}
	\caption{\label{fig:pred_snap}%
		Histograms of the predictive joint distribution $p(x, s_\tau-s)$.
		Noise level $\theta=0, 2$ and speeds $\nu=0.5,1,3$, from top to bottom;
		increasing prediction intervals $\tau$, left to right. Response amplitude
		$A=200$; predictive informations $I=I[s_\tau-s,x]$ as indicated.}
\end{figure*}

Fig.~\ref{fig:pred_snap} shows two-dimensional histograms of output $x(t)$ and
future input change $s_\tau-s$ for various values of the prediction interval
$\tau$. Without noise and for the fast kernel $k_3$, the diagonal shape at low
$\tau$ indicates that the finite-time derivative taken by the kernel, predicts
the future change with good accuracy; slower kernels show this feature to a
weaker extent. As $\tau$ increases, two effects are visible: expansion along
the vertical axis, due to the longer runs for larger $\tau$; and blurring of
the diagonal, which results from the finite length scale of the potential.
(Since we consider virtual runs, tumbling does not contribute to uncertainty
about the future concentration.) At high noise, the diagonal is much weaker,
and the horizontal axis scale is higher; both of these effects result from
noise adding large random contributions to the output. The effect is
particularly strong for the fast kernel.

From histograms as in Fig.~\ref{fig:pred_snap} we estimated mutual information
by taking the sum over bins at index $i,j$
\begin{equation}
	\label{eq:MIestimation}
	I_N^M = \sum_{i,j=1}^{N,N} h(i,j)\log[h(i,j)/h(i)h(j)]
\end{equation}
where $h$ denotes bin counts normalized by the total sample number $M$, or row and
columns counts, respectively. Since this estimator is prone to bias, we used a
procedure~\cite{cheong11a} of subsampling the data to various $M$, and linearly
extrapolating the data points $(1/M,I_N^M)$ to $1/M\to0$. Error bars were
derived from this extrapolation. We verified that the resulting estimate was
consistent for a range of bin numbers $N$.

The resulting predictive information values as a function of the prediction
interval $\tau$ are shown in Fig.~\ref{fig:ivstau} in the main text, where the
predicted variable is the future concentration change. In
Fig.~\ref{fig:icvstau} we show the corresponding result for the predictive
information $I[x, s_\tau]$ between the current output $x(t)$ and the future
input signal itself, $s(t+\tau)$ (rather than the change $s(t+\tau) - s(t)$).
Clearly the predictive information about the future signal value does not
correlate with chemotactic efficiency. The results of Figs.~\ref{fig:ivstau}
and~\ref{fig:icvstau} together show that the search strategy of
\textit{E.\,coli} is based on predicting the future change in the signal,
rather than the signal itself.

\begin{figure}[tb]
	\begin{center}
		\includegraphics[width=1.0\columnwidth]{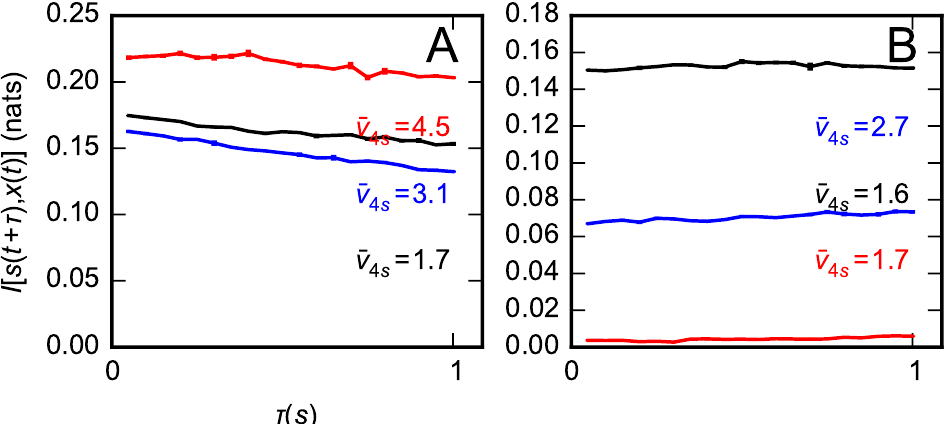}
		\caption{\label{fig:icvstau}%
			Predictive power $I[x, s_\tau]$ between $x(t)$
			and $s(t+\tau)$ \textit{vs.}~forecast interval $\tau$ for
			the original \textit{E.\,coli} kernel with $\nu=1$, for a faster kernel with
			$\nu=3$ and a slower kernel with $\nu=0.5$, for two different noise levels. }
	\end{center}
\end{figure}

} 


\end{document}